\documentclass[12pt,preprint]{aastex}

\shorttitle{Turbulent Outflow}
\shortauthors{Cunningham et al.}

\renewcommand{\vec}[1]{{\bf #1}}

\begin{document}
\title{Protostellar Outflow Evolution in Turbulent Environments}

\author{Andrew J. Cunningham\altaffilmark{1,2}, Adam Frank\altaffilmark{2}, 
Jonathan Carroll\altaffilmark{2}, Eric G. Blackman\altaffilmark{2}, Alice C. Quillen\altaffilmark{2}}
\altaffiltext{1}{Lawrence Livermore National Laboratory, Livermore, CA 94550}
\altaffiltext{2}{Department of Physics and Astronomy, University of Rochester, Rochester, NY 14620}
\email{ajc4@pas.rochester.edu}

\begin{abstract}
The link between turbulence in star formatting environments and
protostellar jets remains controversial.  To explore issues of
turbulence and fossil cavities driven by young stellar outflows we
present a series of numerical simulations tracking the evolution of
transient protostellar jets driven into a turbulent medium.  Our
simulations show both the effect of turbulence on outflow structures
and, conversely, the effect of outflows on the ambient turbulence.  We
demonstrate how turbulence will lead to strong modifications in jet
morphology.  More importantly, we demonstrate that individual
transient outflows have the capacity to re-energize decaying
turbulence. Our simulations support a scenario in which the directed
energy/momentum associated with cavities is randomized as the cavities
are disrupted by dynamical instabilities seeded by the ambient
turbulence.  Consideration of the energy power spectra of the
simulations reveals that the disruption of the cavities powers an
energy cascade consistent with Burgers'-type turbulence and produces a
driving scale-length associated with the cavity propagation length.
We conclude that fossil cavities interacting either with a turbulent
medium or with other cavities have the capacity to sustain or create
turbulent flows in star forming environments.  In the last section we
contrast our work and its conclusions with previous studies which
claim that jets can not be the source of turbulence.
\end{abstract}
\section{Introduction}
Both turbulence and outflows appear to be ubiquitous phenomena
associated with star formation.  Turbulence in molecular clouds is
inferred from Larson's Laws which are empirical relationships between
line-widths and size observed in many star forming regions
\citep{larson}. Turbulence has also been inferred from direct
measurements of power spectra in molecular clouds across a wide range
of scales \citep{heyer}. Though the issue of whether the observed
spectra are a consequence of gravitational fragmentation or mechanical
forcing remains a topic of current debate \citep{field}.  Outflows
driven by newly formed stars also appear uniformly across star forming
environments.  Outflows appear to form very early in the formation of
a star (the Class 0 phase) and continue until a sizable fraction of
the stellar mass has been assembled. In addition, outflows occur
across the entire spectrum of stellar masses from brown dwarfs
\citep{2007IAUS..243..203C, whelan} all the way to high mass O and B
stars \citep{shepherd}.

Given the apparent ubiquity of both turbulence and outflows, one can
ask about their respective importance in terms of impact on the star
formation process.  In recent years the role of turbulence in
moderating star formation has become an active area of research
\citep{maclowkl,krumholz-tr}.  Some workers have argued that turbulent
pressure in the form of a space filling isotropic distribution of
eddies provides support for molecular clouds against gravitational
collapse \citep{goldreich,fleck}.  In this way turbulence is thought
to be the agent which establishes equilibrium and provides a means for
molecular clouds to be long lived structures.  One difficultly with
this view however is the recognition that both hydrodynamic and
magnetohydrodynamic turbulence decays on a timescale of order the
cloud free-fall time $t_{off}$ \citep{stone,maclow-dissipation}.
Thus if clouds are stable, long-lived structures some means must be
found to support them against self-gravitational collapse
\citep{blitz}.

When discussing issues of jet feedback one can distinguish between
three scale lengths.  ``micro-scale'' feedback concerns the effect of
outflows on their own launch scale ($L < 10 AU$).  ``meso-scale''
feedback relates to the impact of jets on environment associated with
infalling envelopes ($L < 10^4~\textrm{AU}$). Considerable work has
focused on these domains: i.e. the way outflows can sculpt
envelopes\citep{delamarter,cunningham-I} or their influence on the
accretion onto the star \citep{krumholz05}. Finally ``macro-scale''
feedback, which is the concern of this paper, is associated with the
effect of outflows on the scale of star forming clusters or the entire
molecular cloud itself.  The role of outflows on large scales became
of interest after it was discovered that some outflows extended to
parsec scales which is of order cloud size or at least the size of
the``clump'' from which star clusters form within a cloud
\citep{1994ApJ...428L..65B,1996ApJ...473..921B,1999A&A...350L..43S, McGroarty}.
Young star clusters such as NGC 1333 are the domains of interest in
this problem.  The impact of many stars, forming roughly co-evaly, on
their parent cloud remains a subject of intense debate and this is
preciously the location where issues in outflows and turbulence
overlap.

Numerous observational studies of star forming regions have shown
that, in general, the energy in active outflows is large enough to
support a cluster or cloud against collapse {\it if that energy can be
coupled to cloud turbulence}
\citep{bally96,ballyreview,knee,quillen,warin}.  These investigations
support the idea that outflows can provide a source of internal
forcing to drive turbulence.  Theoretical treatments of feedback from
multiple spherical outflows creating a self-regulating star forming
system was first explored by \cite{normansilk}.  Analytical work by
Matzner \citep{matzner00,matzner01,matzner} have explored the role of
collimated outflow feedback on clouds. \cite{matzner}, in particular,
developed a theory for outflow driven turbulence in which line-widths
were predicted as functions of a global outflow momentum injection
rate.  \cite{krumholz} have also considered the nature of feedback via
outflows, concluding that these systems provide an important source of
internal driving in dense star forming cores.

The first simulation based study of the problem was presented by
\cite{maclow-outflows} whose results indicated that outflows could
drive turbulent motions.  Decay rates for outflow powered turbulence
in this study were $\sim 4 \times$ faster than that of uniformly
driven turbulence. These early simulations were of relatively low
resolution and further high resolution work is needed.  Recent work by
\cite{linaka} and \cite{nakamura} have mapped out the complex
interplay between star formation outflows and turbulence.  Their
simulations include a self-consistent formulation of driving outflows
from newly formed stars and they concluded that protostellar outflows
were a viable means of generating turbulence in star forming clusters.
However, studies of single jets by \cite{banerjee} came to the
opposite conclusion.  Exploring the volume filling averages of post
shock material they concluded that single jets would not leave enough
supersonic material in their wakes to act as a relevant source of
internal forcing.

Many studies of protostellar turbulence have focused on active
outflows.  These may not, however, be the only means of coupling
outflows with turbulence. Numerical studies \cite{cunningham-jc} have
shown that direct collisions of {\it active outflows} are not
effective at transferring jet/outflow momentum to cloud material (by
active we mean that the momentum flux of the jet or wide angle wind
remains roughly constant through out the interaction). Along similar
lines \cite{quillen} explored the relation between turbulent cloud
motions and observed outflow activity in NGC 1333.  Their results
showed that the location of active outflows did not effect the degree
of turbulence in the cloud.  \cite{quillen} were, however, able to
identify many sub-parsec sized, slowly expanding cavities. These they
identified as dense wind-swept shells of gas that expand into the
cloud after the driving source of the outflow has expired. The net
mechanical energy required to open all these cavities represents a
significant fraction of that required to drive the cloud
turbulence. The authors argued that these cavities were relics of
previously active molecular outflows.  Because cavities have a longer
lifetime than those opened by active and younger outflows, they would
be more numerous than higher velocity, younger outflows. These
cavities could provide the coupling between outflows and turbulence in
molecular clouds. Thus one means of re-energizing turbulence in clouds
may be through``fossil outflow'' cavities which are disrupted after
they have slowed to speeds comparable to the turbulent velocity of the
ambient cloud.  In \cite{cunningham-cavity} high resolution AMR
simulations of transiently driven outflows were carried out.  These
simulations, which included a treatment of molecular chemistry and
cooling, tracked the development of fossil outflow cavities driven by
transient outflows out to near parsec scales.  The simulations
provided observational predictions in terms of position-velocity
diagrams that were consistent with the momentum scaling relations used
in \cite{quillen}.

While the potential for protostars to act as a means of internal
forcing remains attractive, many outstanding questions remain open.
The energetics of turbulent resupply must be understood in terms of a
coupling coefficient between outflow and cloud.  The nature of outflow
interactions must also be specified as \cite{cunningham-jc} have
already shown that active outflows are not efficient it setting large
volumes of gas into random motions. Thus the emphasis shifts to fossil
cavities an their interactions with each other and/or with
pre-existing turbulent medium.  We expect that turbulence can be
generated by collisions between fossil outflows because over the
lifetime of a cluster the cavities will tend to fill the cluster
volume \cite{cunningham-jc}. Individual fossil cavities need not
experience collisions but will still couple to the ambient medium.  With
regard to individual fossil outflow cavities and turbulence the
mechanisms through their directed momentum/energy is``isotropized''
has yet to be articulated.  Something of a chicken and an egg problem
exists in this regard.  In \cite{quillen} expanding outflow cavities
where observed down to speeds a few times that of the turbulence
$v_{outflow} \sim 2 v_{turb} \sim 1~\textrm{km s}^{-1}$. They argued
that the cavities would be subsumed by turbulent eddies once they
decelerated to speeds of order $v_{turb}$.  The expectation that the
directed momentum and energy of a single decelerating cavity can be
given up to turbulence must be explored.  This is the question we seek
to address in this paper.

In the current work we perform a series of a simulations meant to take
the work of \cite{cunningham-cavity} to the next level of
complexity. This paper is meant as an critical intermediary step to
full scale simulations of multiple interacting transient outflows.
Here we consider the role of individual transient jets which are
ejected into a decaying turbulent medium. We note that the evolution
of a radiative jet in a turbulent medium has yet to be explored and
our simulations are relevant to YSO jet morphology and the
reenergization of turbulence via jets and the disruption of the
outflow cavities. Finally our simulations are relevant to issues
addressed in the work of \cite{banerjee} and their conclusions that
jets can not drive supersonic turbulence. We note that the question of
the steady state rate of decay of such flows remains outside of the
scope of the present work as this question could only be addressed by
models that include a steady injection of multiple outflows which are
evolved for longer timescales than the numerical models considered in
this paper.  This issue will be addressed in a forthcoming paper
\citep{carroll}.
\section{Numerical Model}
We have carried forward a series of numerical simulations to
investigate the interaction of young stellar jets with a turbulent
environment.  Each of the simulations is carried forward on a periodic
domain spanning $L_x \times L_y \times L_z = 4000 \times 2000 \times
2000~\textrm{AU}$ discretized to a $504 \times 252 \times 252$
computational grid.  An environment of density
$\rho_a=50~\textrm{cm}^{-3}$ with temperature $T_a=200~\textrm{K}$ is
set into a state of decaying turbulence from an unspecefied source.
Possible candidates for supplying background turbulence within star
forming environments include an initial turbulent state present during
the formation of the cloud from compressive motions in a turbulent
interstellar medium \citep{elmegreen}, differential rotation of
galactic disks, gravitational collapse, the ionizing radiation, winds
and supernovae from massive starts and protostellar outflows
\citep{maclow-dissipation}.  The spectrum of initial motions can be
written as:
\begin{eqnarray}
v_x(\vec{x}) =& \sum_{i,j,k} A_{x~i,j,k} \sin(\vec{k}_{i,j,k} \cdot \vec{x} + \phi_{x~i,j,k}) \nonumber \\
v_y(\vec{x}) =& \sum_{i,j,k} A_{y~i,j,k} \sin(\vec{k}_{i,j,k} \cdot \vec{y} + \phi_{y~i,j,k}) \nonumber \\
v_z(\vec{x}) =& \sum_{i,j,k} A_{z~i,j,k} \sin(\vec{k}_{i,j,k} \cdot \vec{z} + \phi_{z~i,j,k}) \nonumber
\end{eqnarray}
with wave numbers
\[
\vec{k}_{i,j,k} =
 \frac{2\pi i}{2000~\textrm{AU}}~\vec{\hat{x}} + 
 \frac{2\pi j}{2000~\textrm{AU}}~\vec{\hat{y}} +
 \frac{2\pi k}{2000~\textrm{AU}}~\vec{\hat{z}}.
\]
The summation is taken over all $i,~j,~\textrm{and}~k$ between $0$ and
$31$, excluding the $i=j=k=0$ term.  The longest wave-mode seeded in
each spatial direction is therefore equal to the length of the short
edges of the computational grid and the shortest wave-mode in each
spatial direction is resolved across approximately 16 computational
zones.  The phase angles $\phi_{*,i,j,k}$ are chosen at random between
$0$ and $2\pi$. The direction of each wave-mode is determined by
choosing two of the amplitude components, $A_{x~i,j,k}$ and
$A_{y~i,j,k}$ at random between $-1/2$ and $1/2$.  The third amplitude
$A_{z~i,j,k}$ is computed to satisfy the constraint that the initial
velocity field be solenoidal.  Each velocity amplitude
$\vec{A}_{i,j,k}$ is normalized so that the initial velocity power
spectrum follow a power-law dependence on wavenumber
$E_{Ko}(|\vec{k}|) \sim |\vec{k}|^{-\beta}$ with net kinetic energy
prescribed by the desired root mean squared (RMS) turbulent speed
($v_{turb}$) as $V \int E_{Ko}(\vec{k}) d\vec{k} = \frac{1}{2}~\rho_a
v_{turb}$ where $V$ is the volume of the computational domain,
$E_{K}(\vec{k})$ is the power spectrum of velocity perturbations,
\[
E_{K}(\vec{k}) = \left| \mathcal{F}\left( v_x(\vec{x}) \right) \right|^2 +
                 \left| \mathcal{F}\left( v_y(\vec{x}) \right) \right|^2 +
                 \left| \mathcal{F}\left( v_z(\vec{x}) \right) \right|^2,
\]
and $\mathcal{F}$ is the Fourier transform operator,
$\mathcal{F}_{\vec{k}} = \int f_{\vec{x}} e^{i \vec{k} \dot \vec{x}}
d\vec{x}$.  It is also convenient to define the one dimensional power
spectrum as
\[
E_{K}(k) = \int E_{K}(\vec{k}) \delta(|\vec{k}|-k) d\vec{k}.
\]
The simulations which begin with an initially turbulent flow field are
all seeded with an initial RMS turbulent velocity $v_{turb} =
10~\textrm{km s}^{-1}$, yielding an initial RMS turbulent Mach number
of 6.03, and velocity power spectrum index of $\beta=-2$,
characteristic of highly supersonic Burger's-type turbulence
\citep{porter,maclow}.

Our simulations employ periodic boundary conditions at each boundary
interface.  This choice of boundary conditions prevents boundary
artifacts from polluting the spectrum of motions in the grid.
Jet-driven cavities are therefore launched from a grid-embedded region
rather than from grid boundaries.  This choice of domain setup allows
the simultaneous use of periodic grid edges and jet launching.  The jet
launch region extends from $125~\textrm{AU} < x < 250~\textrm{AU}$,
and $(y-1000~\textrm{AU})^2+(z-1000~\textrm{AU})^2 < r_j^2$ where
$r_j= 250~\textrm{AU}$. The launch region maintains a constant density
$\rho_j=250~\textrm{cm}^{-3}$ and temperature $T_j=200~\textrm{K}$
profile and enforces a time dependent jet velocity as
\[
v_x(t) = \left\{ \begin{array}{ll}
  v_o \exp \left(-t/t_{off} \right) & t<2 t_{off} \\
  0 & \textrm{otherwise} \\ 
\end{array} \right.
\]
with initial velocity $v_o=150~\textrm{km s}^{-1}$.  The jet velocity
is smoothed following a quadratic form so that the outer radius of
the launch region have a velocity that is 90\% of the velocity at the
center of the jet.

A total of five simulations have been performed using the AstroBEAR
astrophysical fluid code to carry forward the integration of the Euler
equations using the MUSCL-Hancock shock capturing scheme and Marquina
flux function as described by \cite{cunningham-mhd} with an energy
sink term which models the effects of optically thin radiative energy
loss as described by \citep{cunningham-jc}.  These simulations did not
use the AMR or magnetic field capacities of AstroBEAR.  The suite of
simulations include three models in which jets of with varying decay
parameters are driven into a decaying turbulent environment (Run~1-2),
one model which follows the decay of the turbulent medium in the
absence of disruption by a jet (Run~0) and one model which in which a
non-decaying jet is driven into a quiescent environment (Run~J).  In
table \ref{t1}, we summarize the initial conditions and results each
simulation.  The fifth column from the left tabulates the cooling
efficiency as measured by the net energy loss at the end of the
simulations, the sixth column tabulates the total energy injected into
the grid at the jet launch region over the duration of the simulations
relative to the initial turbulent energy seeded in the grid and the
rightmost column tabulates the least squared fit to the velocity spectrum
power law index at the end of the simulation $\beta |_{t=tf}$ for
$10^{1.5}~\textrm{AU}<k<10^{2.3}~\textrm{AU}$.  In our simulations the
power-law spectral dependence expected for inviscid turbulent flow
begins to break for larger k due to numerical dissipation and for
smaller k due to the inability to represent modes with characteristic
length greater than the shorter edges of the simulation domain.
\begin{table}[!h] \caption{Simulation Summary.\label{t1}}
 \begin{tabular}{l | c c c c c c}
   & $t_{off}$ & Turbulence & Final Time $t_f$ & Cooling Efficiency & $\frac{E_{K,jet}}{E_{Ko}}$ & $\beta |_{t=t_f}$\\
 \tableline
 Run~0 & no jet           & $\checkmark$ & 922 yr & 36.1\% & -     & -2.09 \\
 Run~1 & 8 yr             & $\checkmark$ & 922 yr & 49.0\% & 0.285 & -2.24 \\
 Run~2 & 100 yr           & $\checkmark$ & 221 yr & 9.45\% & 3.56  & -2.38 \\
 Run~3 & non-decaying jet & $\checkmark$ & 161 yr & 22.7\% & 71.0  & -2.43 \\
 Run~J & non-decaying jet & -            & 161 yr & 22.1\% & -     & -2.46 \\
 \end{tabular}
\end{table}

The simulations presented in this work are of considerably shorter
timescales than that of actual protostellar outflows.  This is
primarily due to computational constraints imposed by the need to
simultaneously resolve the overall outflow, the outflow injection
radius $R_j$ and turbulent eddies with scale length $< R_j$.  The
simulations of extinct outflow sources presented here are interpreted
as models of the fossil relics of extinct protostellar outflow
activity in molecular clouds.  The simulated of fossil cavities are
intended to be roughly comparable to actual protostellar outflow
cavities at comparable time relative to the outflow shut-off time,
$t/t_{off}$.
\section{Results}
\begin{figure}[!h]
\begin{center}
\begin{tabular}{l}
  \includegraphics[angle=-90,clip=true,width=0.3\textwidth]{f1a.ps}
  \includegraphics[angle=-90,clip=true,width=0.45\textwidth]{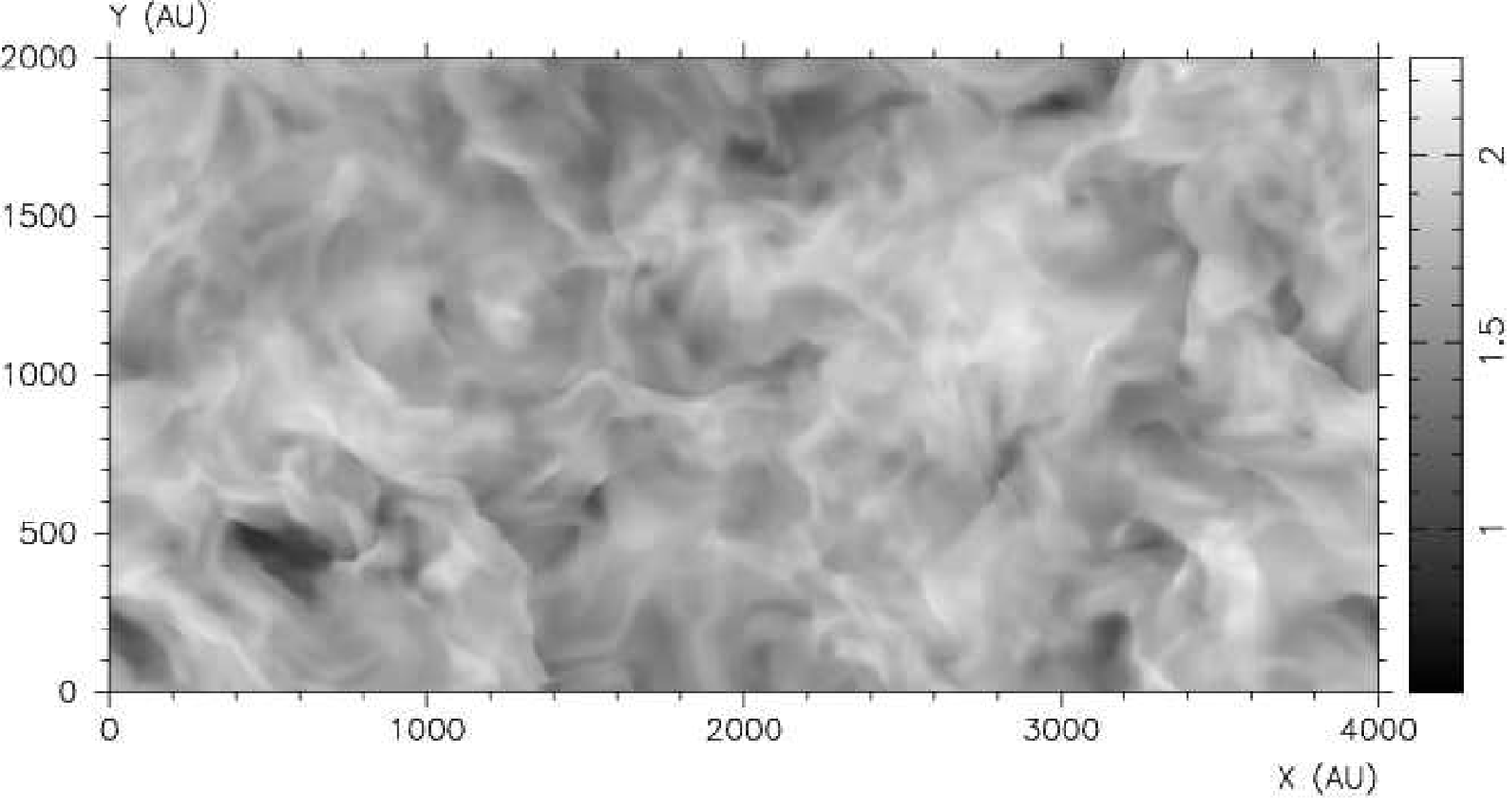} \\
  \hline
  \includegraphics[angle=-90,clip=true,width=0.3\textwidth]{f1c.ps}
  \includegraphics[angle=-90,clip=true,width=0.45\textwidth]{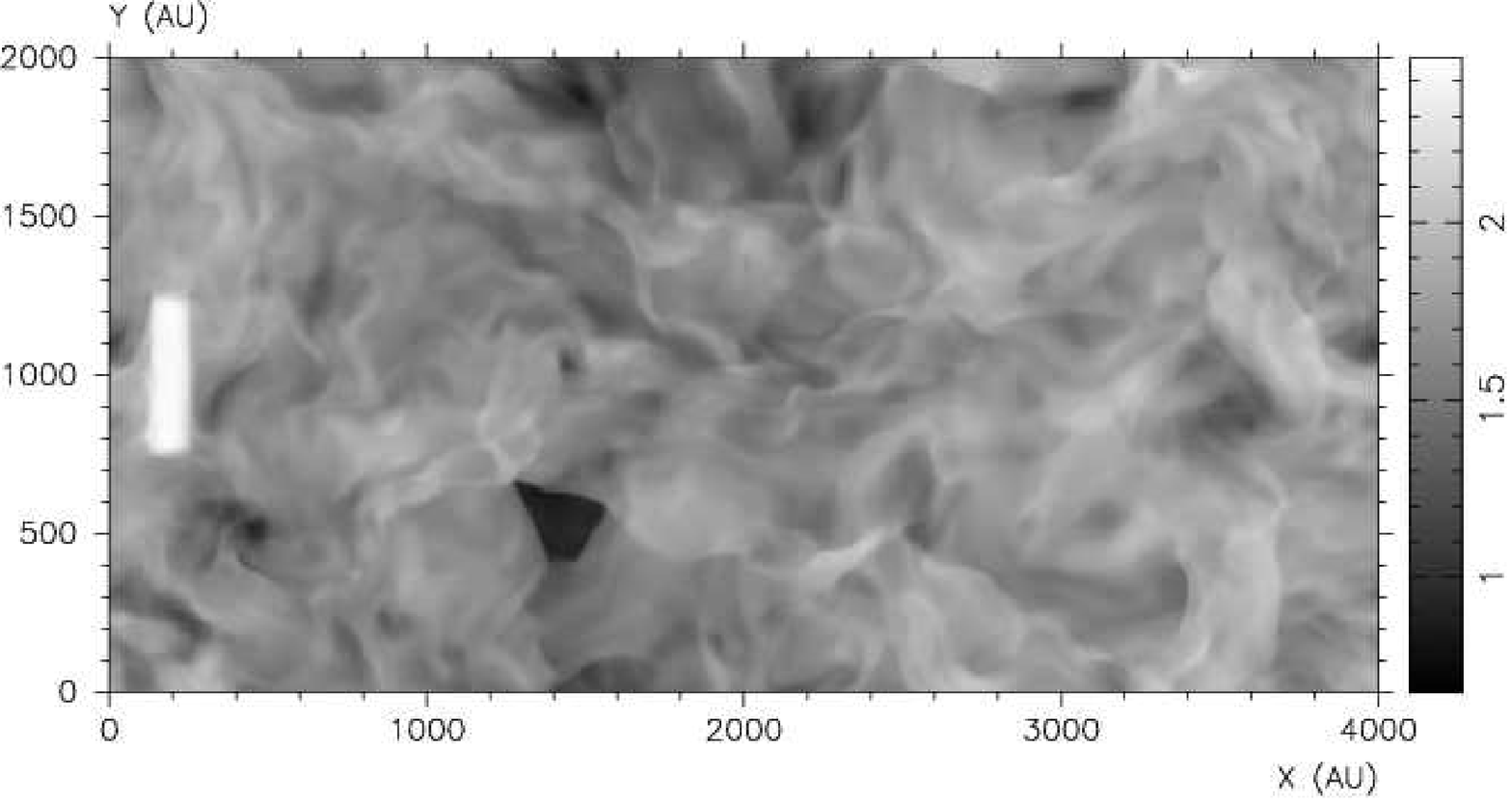} \\
  \hline
  \includegraphics[angle=-90,clip=true,width=0.3\textwidth]{f1e.ps}
  \includegraphics[angle=-90,clip=true,width=0.45\textwidth]{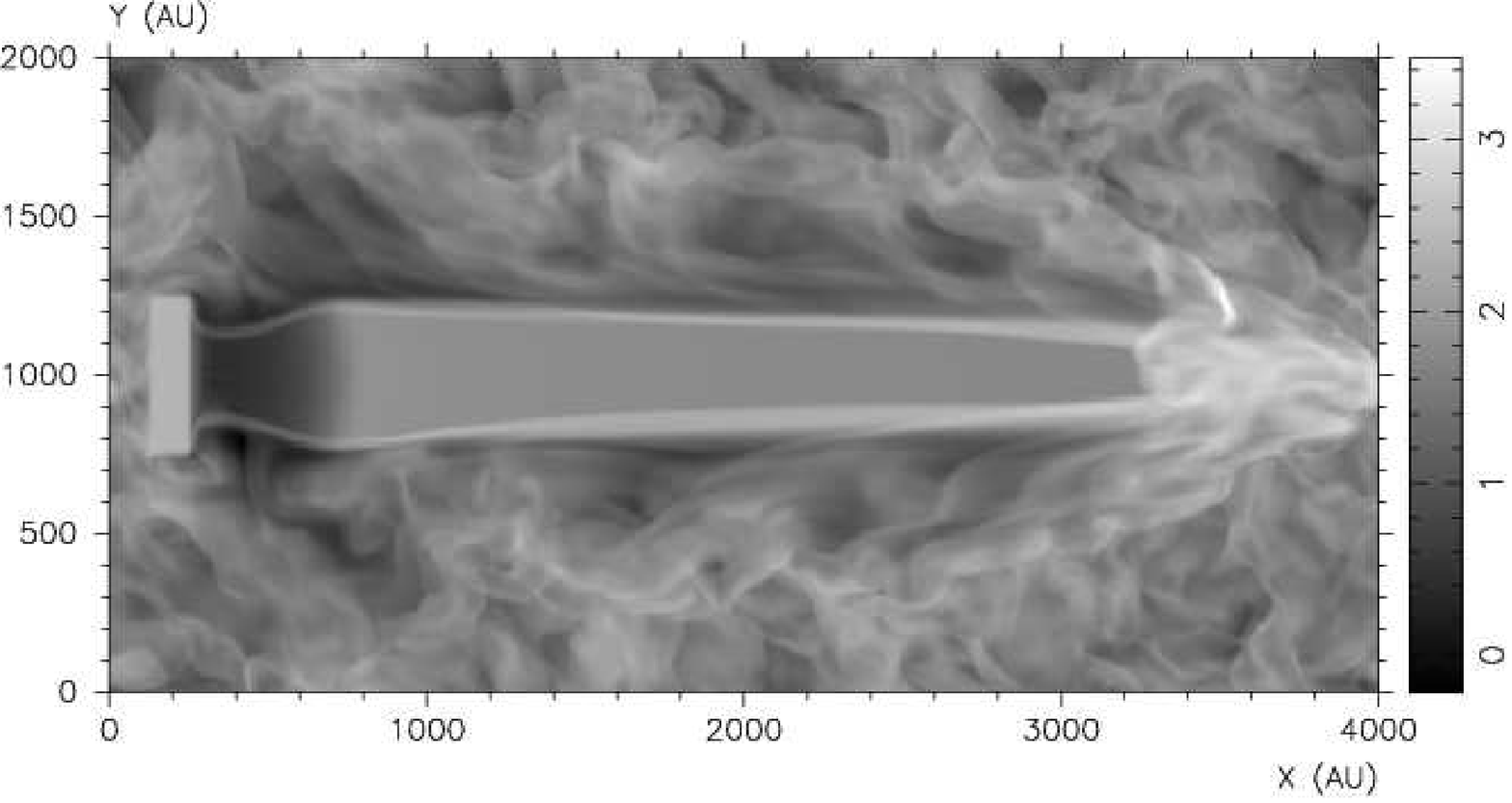} \\
  \hline
  \includegraphics[angle=-90,clip=true,width=0.3\textwidth]{f1g.ps}
  \includegraphics[angle=-90,clip=true,width=0.45\textwidth]{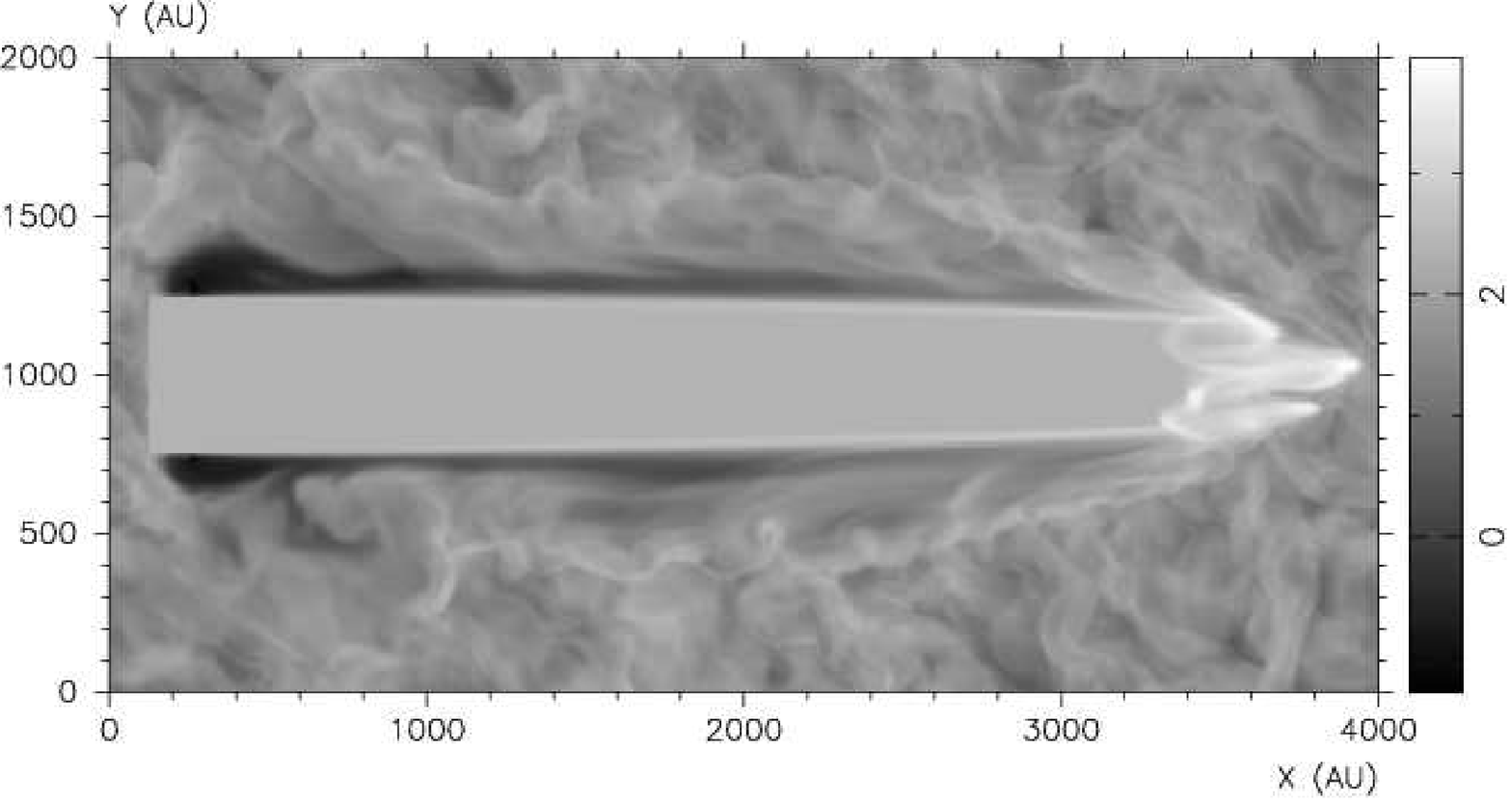} \\
  \hline
  \includegraphics[angle=-90,clip=true,width=0.3\textwidth]{f1i.ps}
  \includegraphics[angle=-90,clip=true,width=0.45\textwidth]{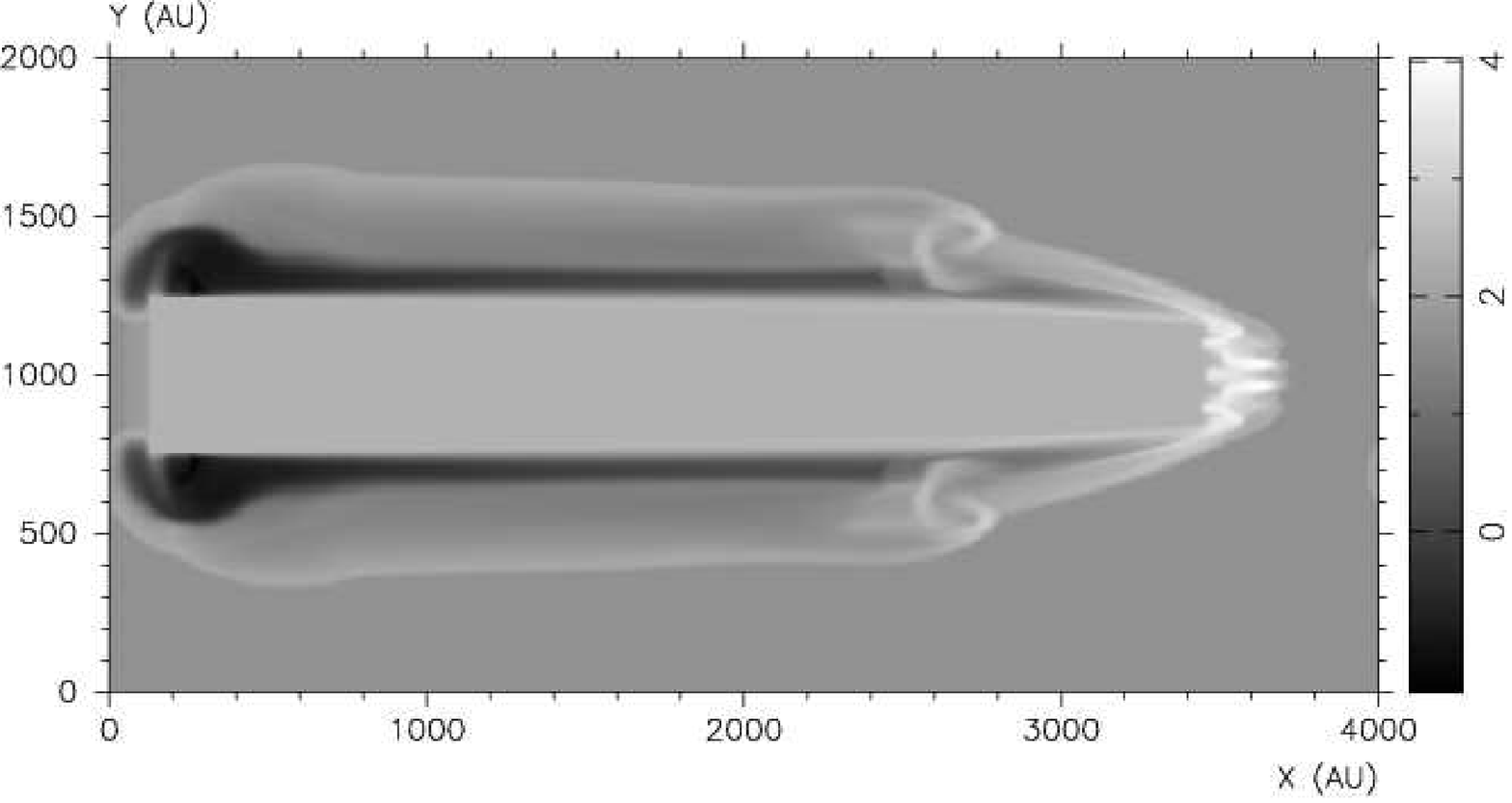} \\
\end{tabular}
\end{center}
\caption{Velocity power spectra at several epochs (left) and
crosscuts about the mid-plane showing the logarithm of the gas
density, $\log(\rho [\textrm{cm}^{-3}])$, at the end of the each
simulation.  The panels show Run0, Run1, Run2, Run3, and RunJ from top
to bottom. \label{f1}}
\end{figure}

The right panel of figure \ref{f1} shows a crosscut about the
mid-plane of the simulation domain at the end ($t=t_f$) of each
simulation.  The initial and jet launch conditions of each simulation
are summarized in the left three columns of table \ref{t1}.  The right
three columns of table \ref{t1} list diagnostic parameters at the end
of each simulation.  The ``cooling efficiency'' which is taken as the
fraction of the total energy budget lost from the simulation domain
via radiative cooling is listed for each run.  This quantity measures
the radiative loss of energy from the system that may have otherwise
been available to sustain turbulent motion.  Note that this cooling
efficiency increases for simulations that run for longer physical
time.  This is because these simulations have more time for radiative
energy loss to occur.  It also increases for simulations with longer
jet decay times as these jets feed energy into their bow shocks for
longer times and have stronger cooling compared with jets that turn
off earlier.

The propagation speed of the Mach disk structure at the head of the
simulated outflows can be estimated analytically for the case of the
continuously driven jet models (run~3 and run~J).  \cite{blondin}
consider the momentum flux across the inward and outward facing
surfaces of the Mach disc structure at the head of jet-driven flows to
derive an analytic estimate for jet bow-shock propagation speeds,
\[
v_{bs}=v_j \left[1+\left(\frac{\rho_j}{\rho_a}\right)^{-1/2}\right]^{-1}.
\]
This expression evaluates to $v_{bs}=104~\textrm{km s}^{-1}$ for the
continuously driven jet model parameters considered in this paper.
The shocks delineating the Mach disk at the head of the continuously
driven jets (run~3 and run~J) propagate $\sim3500~\textrm{AU}$ over
$t_f=161~\textrm{yr}$, corresponding to propagation speeds of
$103~\textrm{km s}^{-1}$, in excellent agreement with the analytic
estimate.  Comparison of the two simulations shows that the turbulence
does not effect the average propagation speed of driven jets.

\begin{figure}[!h]
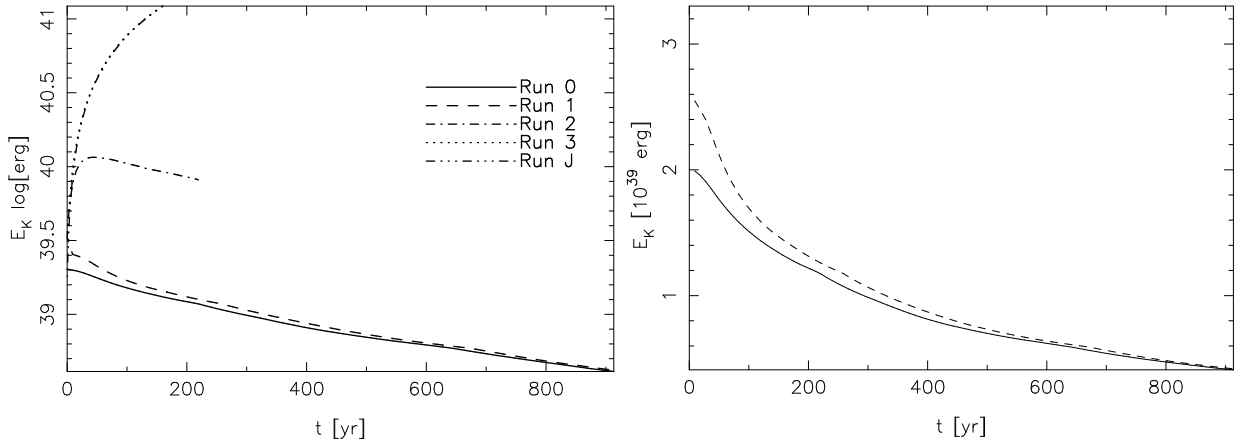

\begin{center}
  \includegraphics[angle=-90,clip=true,width=0.49\textwidth]{f2a.ps}
  \includegraphics[angle=-90,clip=true,width=0.49\textwidth]{f2b.ps}
\end{center}
\caption{Evolution of the total kinetic energy in the computational
domain for each simulation (left) and the evolution of the kinetic
energy for run0 and run1 only (right). The kinetic energy of both
run~J and run~3 are dominated by the continuously driven jet and the
two curves are indistinguishable in the left plot.  Note the similar
form of decay for the jet with turbulence cases (run~1 and run~2) to
that of the turbulence-only control (run~0).  \label{f2}}
\end{figure}

Figure \ref{f2} shows the time evolution of the total mechanical
energy in the grid for each simulation.  Because the mechanical energy
of the continuously driven jet quickly overwhelms that of the
turbulent environment, the time evolution of the mechanical energy for
run~3 and run~J are indistinguishable in figure \ref{f2}.  For
decaying jets, run~1 and run~2, the net mechanical energy increases
until the driving source begins to decay at $t \sim t_{off}$.
Interestingly, for $t > t_{off}$ the mechanical energy decays in a
roughly similar form as that of the control simulation, run~0, which
which follows the decay of the turbulent environment alone 

%
\clearpage
\subsection{Jet Morphology}
\begin{figure}[!h]
\begin{center}
  \includegraphics[angle=-90,clip=true,width=0.49\textwidth]{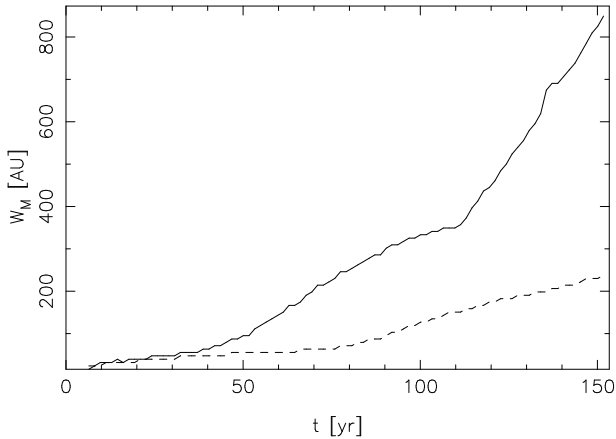}
\end{center}
\caption{Temporal evolution of the Mach Disk height of the
continuously driven jet in a turbulent environment, run~3 (solid line)
and the continuously driven jet in a quiescent environment, run~J
(dashed line) plotted on a linear-linear scale. \label{f22}}
\end{figure}

The morphological effects of the turbulent medium on the jet-driven
bow-shock in run 3 is readily apparent despite the comparatively slow
turbulent speed in the ambient medium ($v_{turb}$ is only 10\% of the
jet propagation speed).  This is particularly apparent in the change
in behavior of the vortex ring in run~J located near $x \sim
2750~\textrm{AU}$ in comparison to run~3.  Specifically the formation
of a clearly defined vortex ring is suppressed for the case of the
identical jet conditions which are driven into a turbulent medium in
run~3.  Furthermore, the entire bow shock from the base of the jet to
its head shows significant disruption compared to the control case.
This is most apparent, and can be most easily analyzed quantitatively,
at the Mach disk at the jet head.  In the jet which is driven into a
quiescent environment (run~J) a thin shock-bounded Mach disk structure
forms at the head of the outflow.  In the case where the same jet
launch conditions are imposed into a turbulent environment (run~3),
the Mach disk is observed to broaden significantly.  Figure
\ref{f22} presents a comparison of the Mach disk width ($W_M$) for
both runs.  The figure was constructed by measuring the extent of the
Mach disk at its widest point using compression in the jet head as a
tracer ($\rho > 750~\textrm{cm}^{-3}$).  The figure shows the Mach
disk width increasing in the turbulent simulation by more than a
factor approximately 4 over that in the control case.

To quantify this change run~J shows a Mach disk width of $W_M \sim
250\textrm{AU}$.  Dividing by the age of the jet yields an average
expansion rate of the Mach disk of $v_{exp}=W_M/t_f=7.1~\textrm{km
s}^{-1}$.  For run~3, the jet driven into the turbulent environment,
the Mach disk structure shows $W_M \sim 800\textrm{AU}$.  The average
expansion rate is $v_{exp}=W_M/t_f=22.7~\textrm{km s}^{-1}$.  Note
that the additional expansion of the Mach disk in the turbulent case
exceeds the initial ambient turbulent velocity,
$v_{turb}=10~\textrm{km s}^{-1}$.  Thus it appears that more than just
the entrainment of ambient turbulent motion is at play in this
behavior.

The change in the behavior of Mach disk highlights the fact that even
structures which are denser and propagate more quickly than the
ambient turbulence are significantly disrupted in its presence.  This
result suggests an interpretation of jet evolution in a turbulent
medium which can account for the eventual disruption of fossil
cavities in decaying jets.  It is well known that initially laminar
shock flows can be disrupted via the action of various instabilities
associated with strong cooling: i.e. radiative \citep{sutherland} and
thin shell modes \cite{vishniac1,vishniac2}.  Given this susceptibility
to fluid instability, the motion of the ambient turbulence with eddies
occurring at a variety of scales should provide a space filling
environment of multi-mode perturbation seeds.  Because the turbulent
field is composed of a continuous spectrum of eddies it will seed the
outflow shell with perturbations which correspond to the fastest
growing modes of any instability which the outflow-swept shell may be
subject.  These seeds act to initiate instability growth in the bow
shock delineating the cavity and facilitate the coupling of the energy
and momentum of the original outflow to the turbulent environment.

Consider for example a shell of shocked gas of width $h$ with internal
sound speed $c$.  The growth rate for modes of the Non-Linear Thin
Shell instability for an initial displacement $L$ can be approximated
as \cite{hueckstaedt}:
\begin{equation}
\Gamma \sim C_d^{-1/2} c k\sqrt{k L}
\end{equation}
Where $C_d$ is of order 1.  The growth of the instability requires the
displacement $L$ be of order the shell width $L \sim h$. Since the shells
are thin $h << r_j$ and the time scale for the bow-shock to fragment
will be short compared to a characteristic hydrodynamic timescale of the
flow which in this case we could take to be the crossing time of the
jet across the cluster scale $t_{cl} \sim r_{cl}/v_j$.  Thus considering
only the non-linear thin shell instability with $L \sim h$ and $k \sim
1/h$ we see that 
\begin{equation}
\Gamma t_{cl} \sim \frac{c}{v_j} \frac{r_{cl}}{h} \sim M^{-1} \frac{r_{cl}}{h}.
\end{equation}
Given typical Mach numbers in YSO jets of $M \sim 100$ and $r_{cl}
\sim 1~\textrm{pc}$ we have $\Gamma t_{cl} >> 1$. Thus perturbation
seeds provided by the turbulence will lead to the fragmentation of the
bow shock by the time the cavity has reached is maximum extent.

\begin{figure}[!h]
\begin{center}
  \includegraphics[clip=true,width=0.75\textwidth]{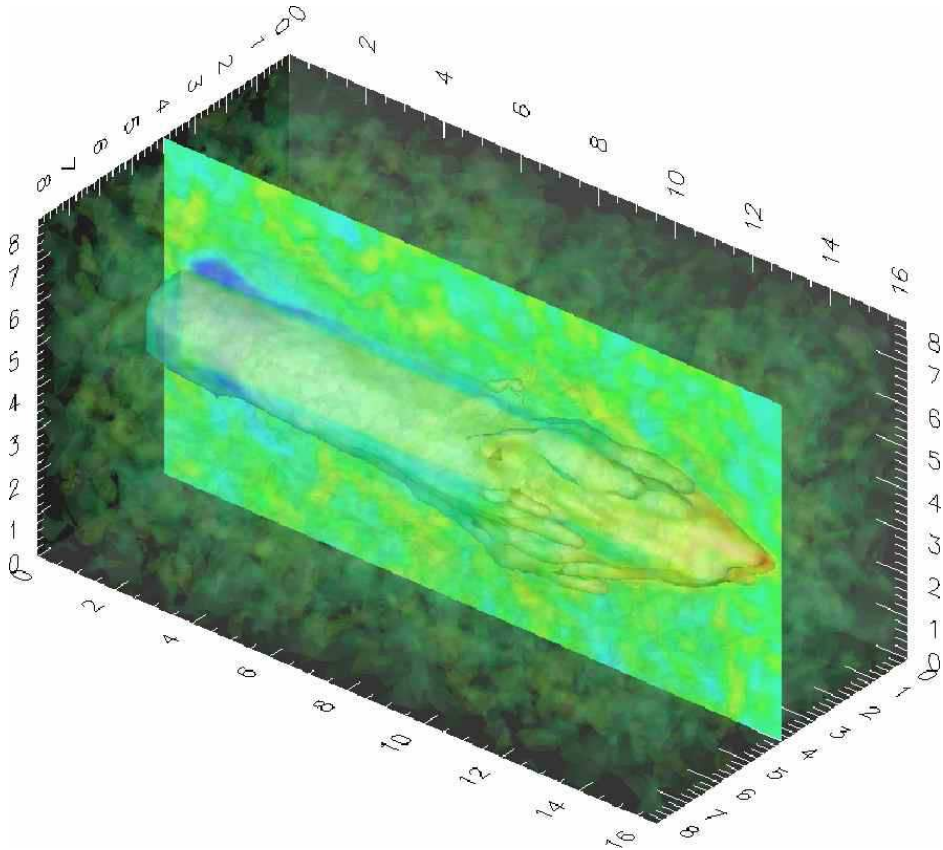}
\end{center}
\caption{Three dimensional volume rendered realization of the
logarithm of the gas density, for the continuously driven jet in a
turbulent environment with semi-transparent isosurfaces rendered at
$\rho = {2.33,~3.67,~\textrm{and}~316}~\textrm{cm}^{-3}$. \label{f3}}
\end{figure}

Putting the issue of the generation of turbulence aside for a moment
we see that issues associated with the propagation of active jets are
raised when we consider the disruption of the bow shock.  Figure
\ref{f3} shows a three dimensional visualization of run~3. Note the
fragmentation of the outflow into clumpy filaments. In particular,
note the fragmentation of the Mach disk associated with the jet
bow-shock complex into clumps.  Each clump drives a smaller scale
bow-shock in its wake, resulting in the formation of filamentary
density enhancements in the Mach disk.  Similar irregular structures,
including clumps and filaments in the Mach disk, have been observed in
a number of jet systems including HH~1/2 and HH~47 as has been
discussed by \cite{bally} and \cite{hartigan} respectively.
\cite{yirak} have noted similar bow-shock structures when the head of
a propagating jet is impeded by stationary density homogeneities
embedded within the ambient environment.  Thus it appears that
filamentary fragmentation of protostellar jet bow-shocks may be
characteristic of an inhomogeneous environment in general, regardless
of the form of such inhomogeneities.

Observations by \cite{2003AJ....126..893B} and
\cite{2006AJ....131..473B} also reveal shock delineated knotty
deflected structures within protostellar jet beams.  Such deflections
are puzzling when they appear discontinuous, with the jet showing a
sharp deflection only some ways down the beam.  While \cite{yirak}
find that sufficiently dense inhomogeneities can cause similar jet
deflections, the motions of the turbulent environment in the present
simulations show no comparable effect on the continuously (run~3) or
long-driven (run~2) jets.  However, the turbulent motions of the
ambient environment do sufficiently disrupt the lateral edges of the
jet bow-shock structure to expose the driving jet beam to the
influence of nearby turbulent eddies.  It is interesting to note that
the turbulent eddy scales which are represented in the computational
domain $(L \le 4000~\textrm{AU})$ do act to compress the edges of the
long decay time jet beam (run~2).  Thus it is possible that
supra-jet-scale flow eddies ($L \gg \textrm{bow-shock radius}$) which
would have higher velocity may drive bending and global disruption of
jet structures.  On the issue of large-scale disruption, We note the
work of \cite{lebedev} in which the deflection of supersonic jets via
interaction with a large-scale transverse flow was considered.  This
study explored the issue both experimentally and numerically and
addressed the issue of a jet interacting with side-wind after
propagating some distance from the jet source.  These studies raise
the possibility of an encounter with a large eddy as the cause of
discontinuous jet bending seen in observations.  However, the effect
of very large scale eddies are not considered in the present study as
their inclusion would require a much larger computational volume and
many more grid points to simultaneously achieve adequate resolution of
the jet and the large scale flow.
\subsection{Turbulence and Kinetic Energy Power Spectra}
The left column of figure \ref{f1} shows plots of the velocity power
spectra for each simulation at several times over the course of the
simulations.  The power spectra have been constructed by binning the
three dimensional Fourier transform of the computational domain into
spherical slices of spectral width $\Delta k = 2 \pi / \Delta x =
0.792~\textrm{AU}^{-1}$.

First we note the degree to which the outflows interact with the
background turbulence. Only the jet driven cavity with the shortest
driving decay time in run~1 becomes completely disrupted by, and
subsumed into, the background turbulence.  However, the more slowly
propagating radial edges of the cavity bow-shock in the longer jet
decay time runs all show some degree of coupling with the ambient
turbulence.  In the case of the continuously driven jet in the
turbulent environment (run~3), approximately one cavity bow-shock
radius ($R_{BS}$) crossing time $(t_{BS} \approx R_{BS}/v_{turb})$ has
elapsed by the end of the calculation. In that simulation the oldest
portion of the outflow limb, which is closest to the outflow source,
shows the most disruption qualitatively.

Moving now to the power-spectra we see the longer-lived jet sources of
run~2 and run~3 show increasing power in the largest flow eddies
(smallest wavenumber).  This indicates that the outflow cavities in
these simulations provide sufficient power at length scales comparable
to the width of the simulation domain to support turbulent motions in
the ambient flow against energy decay.  Even though the rapidly
decaying jet (run~1) injects comparatively less energy, we also find
more energy in the largest eddies compared to the case of turbulence
without a jet.  Much of the mechanical energy injected into the domain
by the outflow source appears as large scale motion (smallest
wavenumber) in this simulation as well.

The increasing power at small wavenumber exhibited by the jet models
can be interpreted as the net result of the decay of turbulent energy
at large wavenumber along with the resupply of turbulent energy by the
injected outflow toward smaller wavenumber.  The more powerful outflow
source in run~3 resupplies energy toward small wavenumbers faster than
the turbulent cascade of drains energy toward larger wavenumbers.  In
the case of the outflow cavity models (run~1 and run~2), modes with
the smallest wavenumber increase even as the total kinetic energy in
the grid decreases (figure \ref{f2}).  We therefore {\it identify the
driving scale of turbulent resupply} associated with the scale of
cavity propagation as $k_{drive} \sim 1/L_{BS}$.  Furthermore, the
simulations that drive outflows exhibit steeper power-law slopes than
the decaying turbulence case.  At the end of the respective
simulations, the decaying turbulence model in run~0 reveals a spectrum
with power law index $\beta=-2.09$ compared to the jet-only model with
$\beta=-2.46$.  These cases bracket the slope of the spectra for the
outflow with ambient turbulence models which show a clear steepening
with increasing outflow energy (Table \ref{t1}).  This steepening can
be interpreted as an outflow-induced suppression of sub-cavity scale
flow eddies.  The power spectra therefore indicate that outflow
cavities act to support the ambient turbulence in two ways.  First the
outflows power eddies of comparable extent to the length of the cavity
$L_{BS}$ and second, the opening of outflow cavities inhibit the
cascade of energy to sub-outflow scales.

We now address the issue of cavity disruption. Note that density
distribution of run~3 in the neighborhood of the jet is characteristic
of a uniform, laminar flow.  After a short time, the energy injected
by the jet greatly exceeds that of the ambient turbulence.  Therefore,
the power spectra at the end of this simulation is dominated by the
jet at all scales and the spectra of run~3 appears indistinguishable
from that of the jet-only control in run~J.  The density distribution
crosscuts shown in the right column of figure \ref{f1}, however,
reveal that the outflow cavity driven by the short, nearly impulsive
jet in run~1 has been completely subsumed into turbulent eddies.  By
the end of the simulation the flow field for the short jet pulse
(run~1) is qualitatively similar to that of the turbulence-only
control (run~0).  This is the behavior that was expected by
\cite{quillen} and \cite{cunningham-cavity} for long extinct outflow
structures embedded in turbulence.  Furthermore, the net mechanical
energy of the disrupted outflow cavity in run~1 decays at a rate
comparable to that of the turbulence-only control in run~0.  This
implies that the disrupted outflow cavity becomes turbulent itself as
it evolves under the influence of the turbulent environment.

\begin{figure}[!h]
\begin{center}
  \includegraphics[angle=-90,clip=true,width=0.32\textwidth]{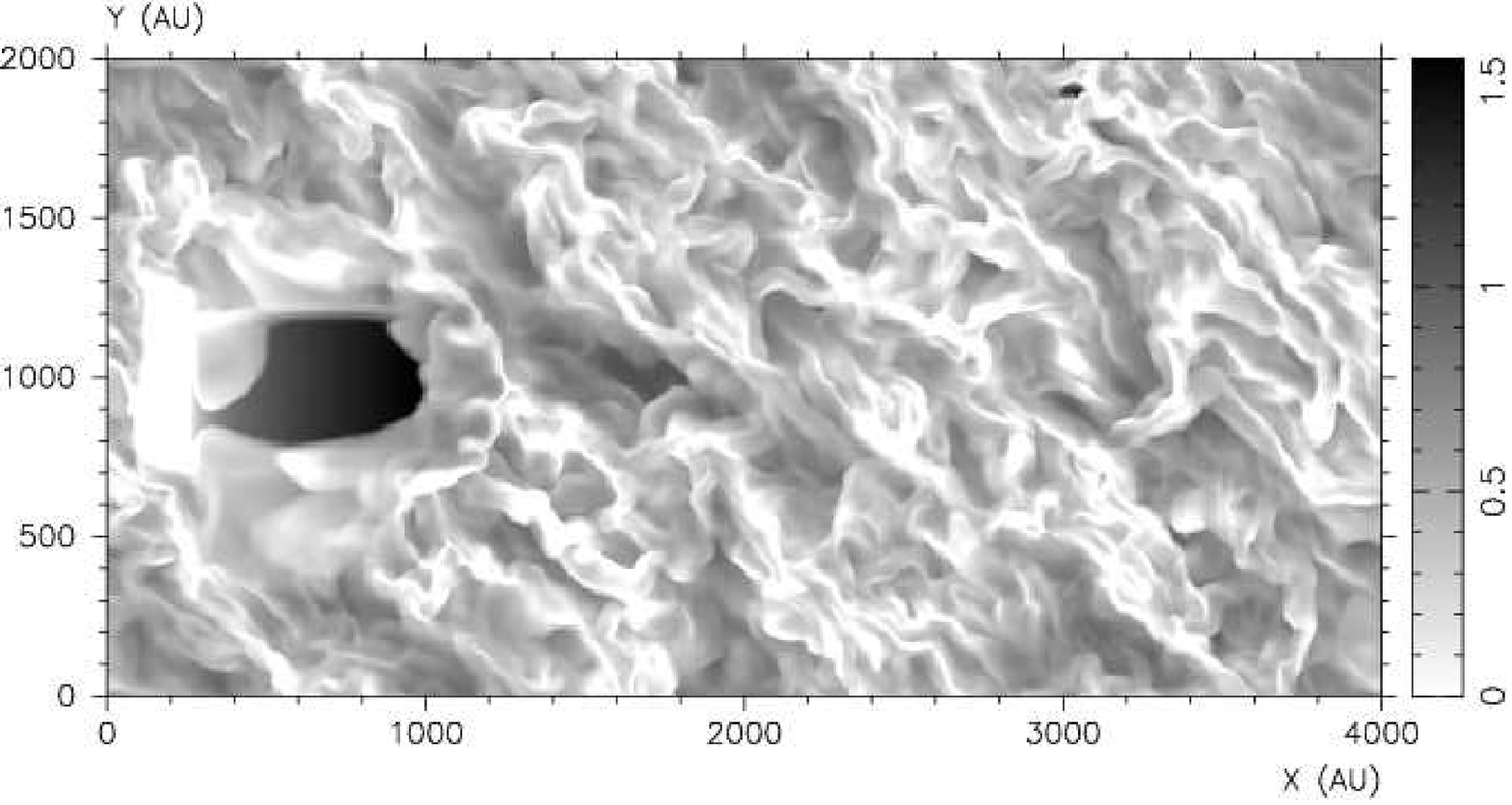}
  \includegraphics[angle=-90,clip=true,width=0.32\textwidth]{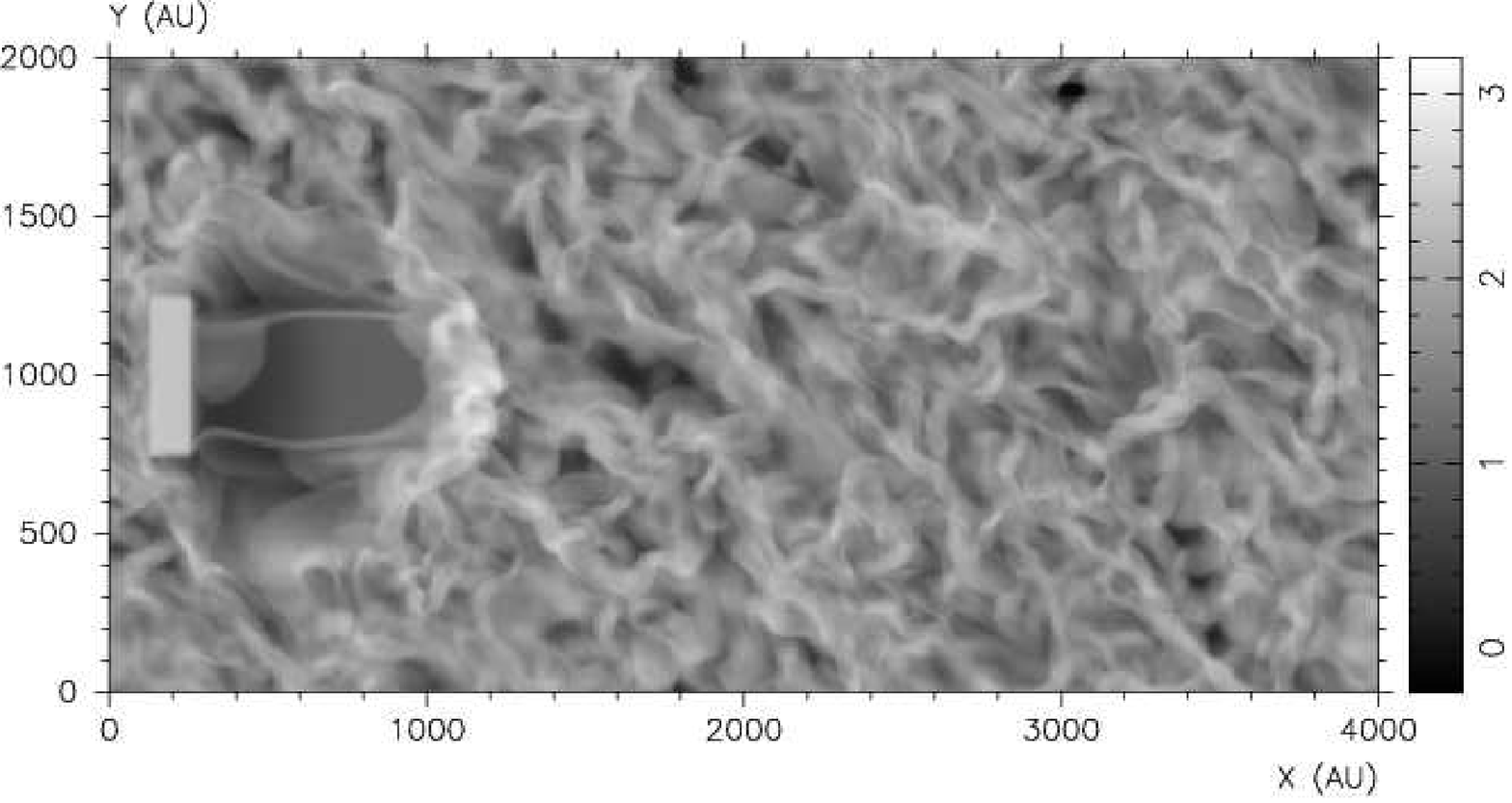}
  \includegraphics[angle=-90,clip=true,width=0.32\textwidth]{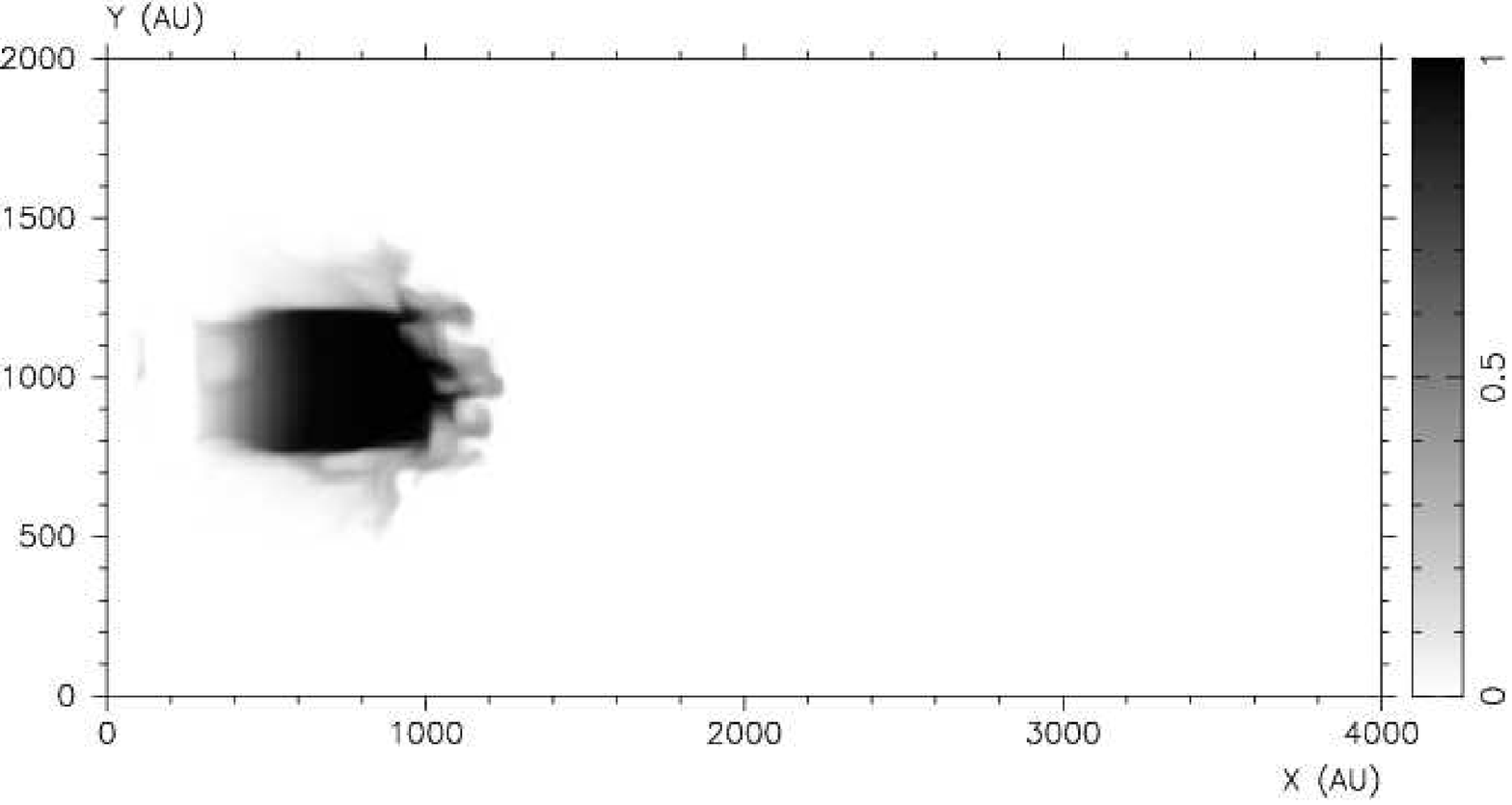} \\
  \includegraphics[angle=-90,clip=true,width=0.32\textwidth]{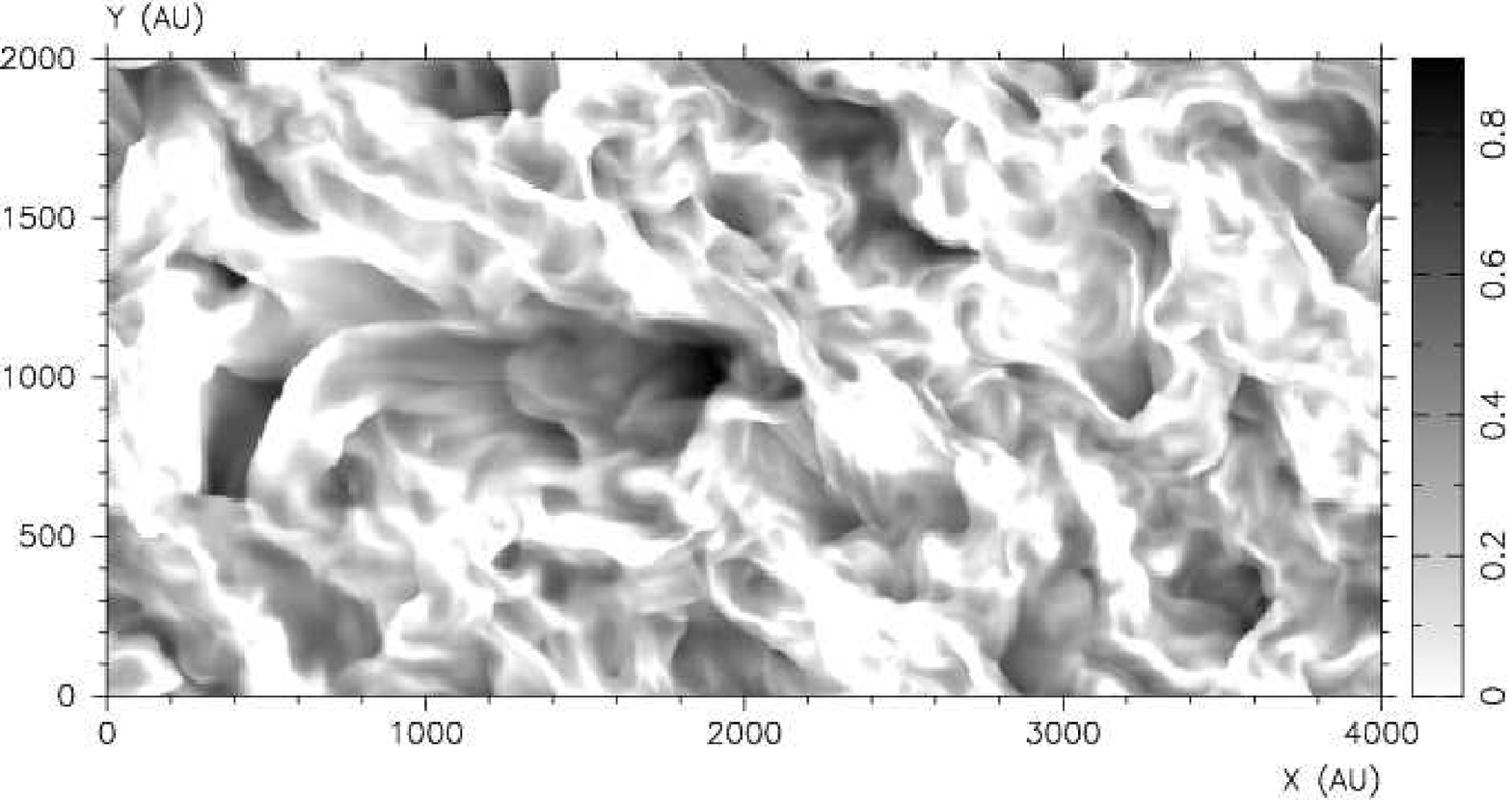}
  \includegraphics[angle=-90,clip=true,width=0.32\textwidth]{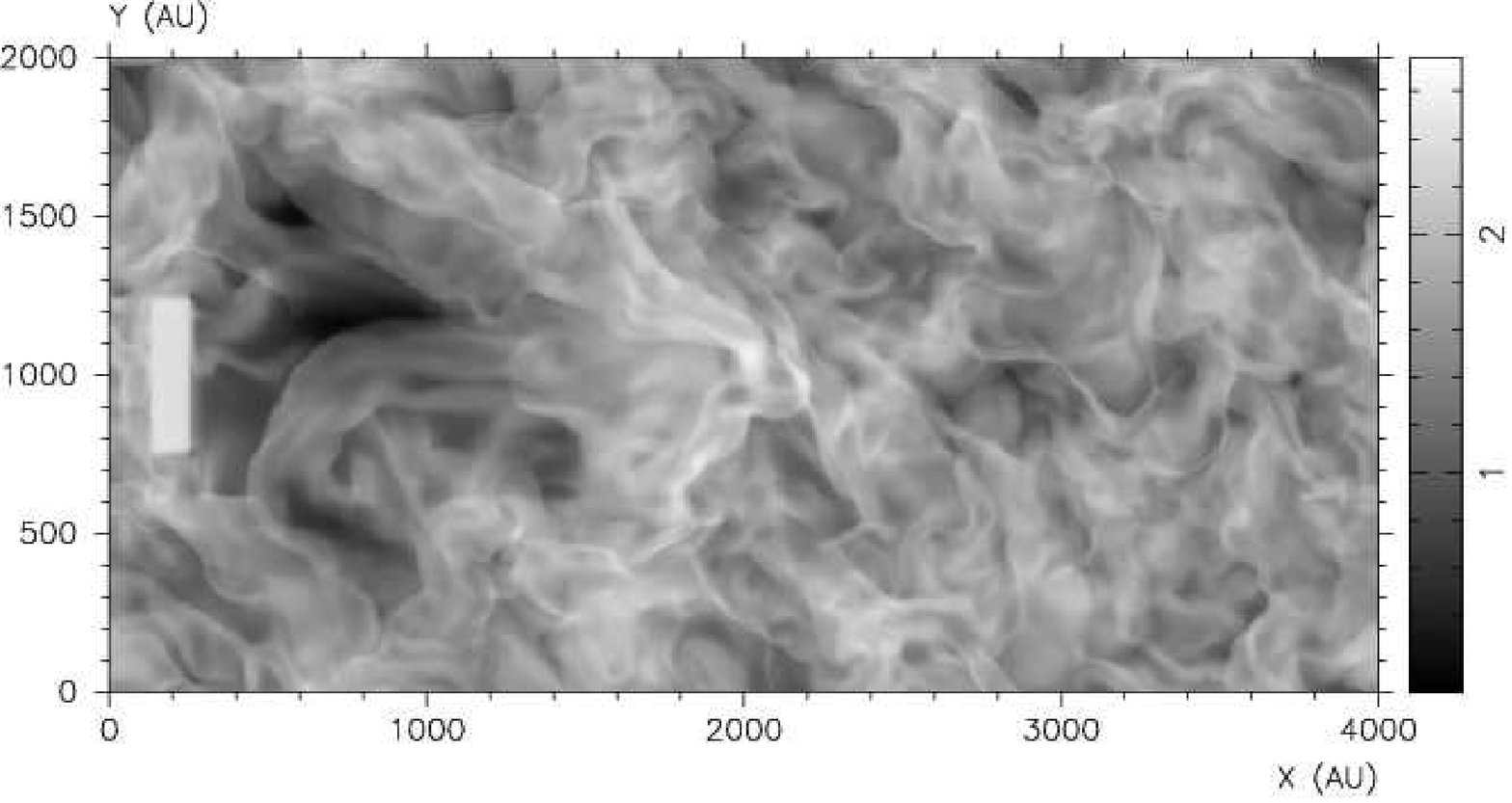}
  \includegraphics[angle=-90,clip=true,width=0.32\textwidth]{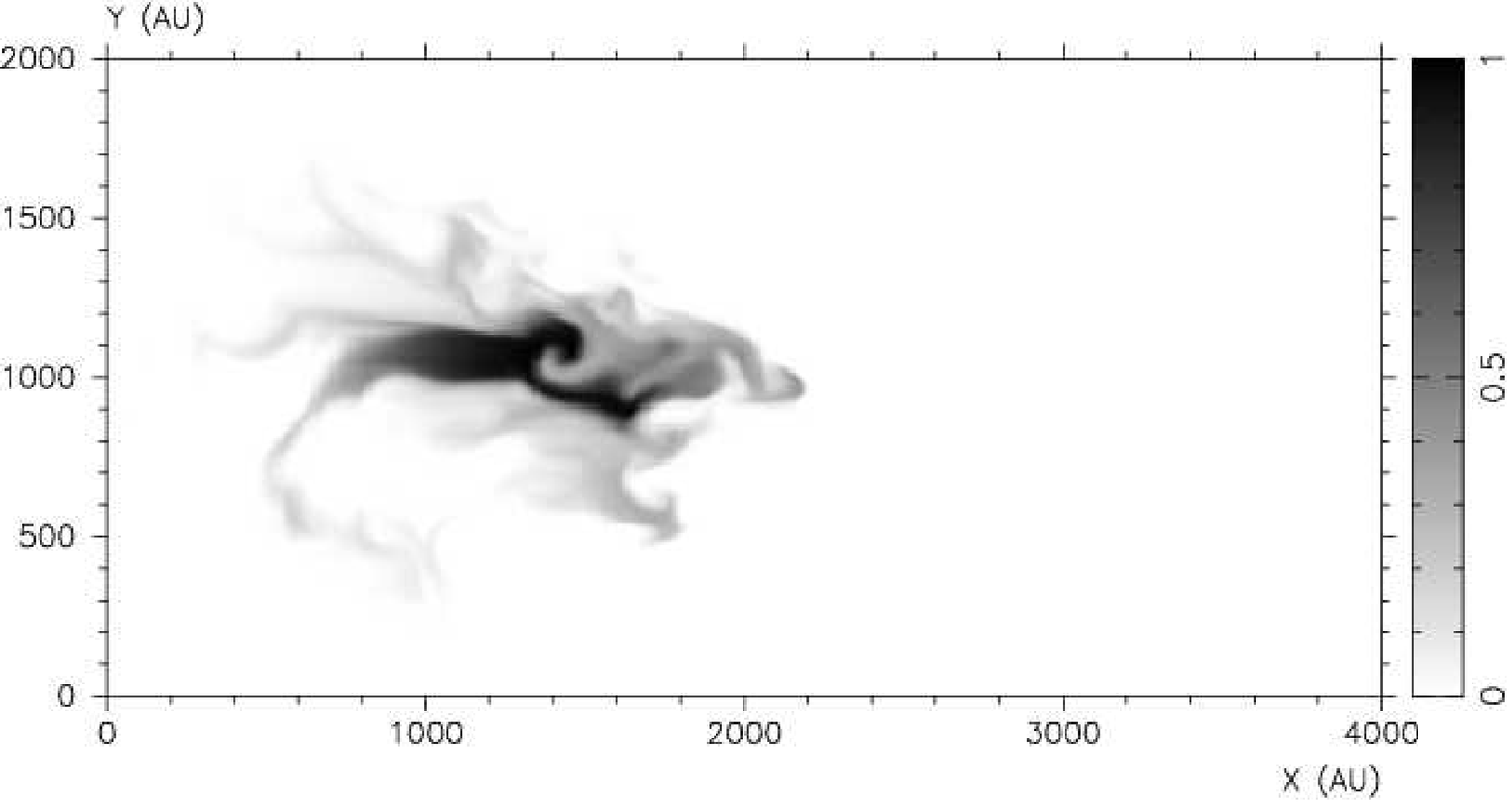} \\
  \includegraphics[angle=-90,clip=true,width=0.32\textwidth]{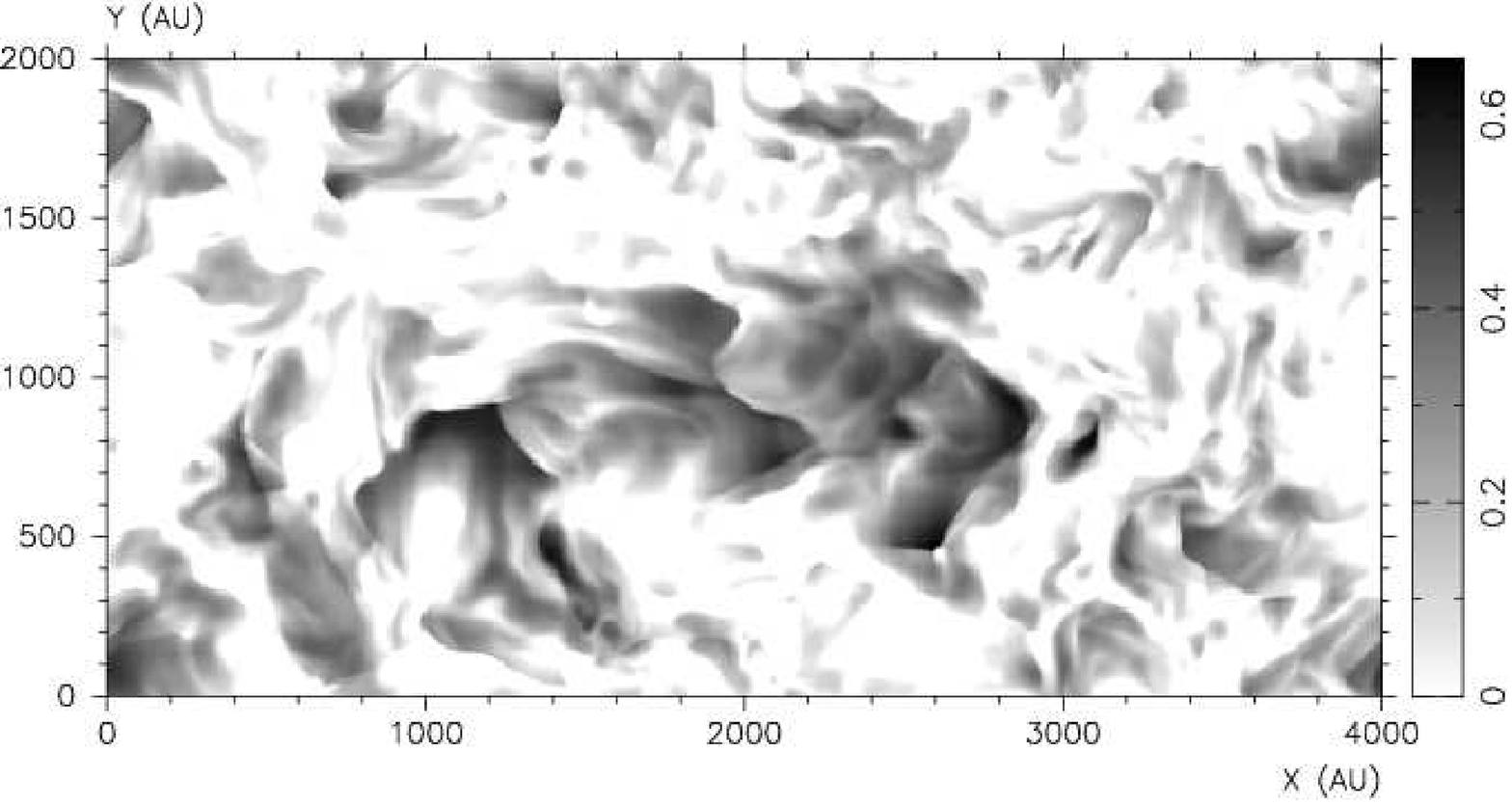}
  \includegraphics[angle=-90,clip=true,width=0.32\textwidth]{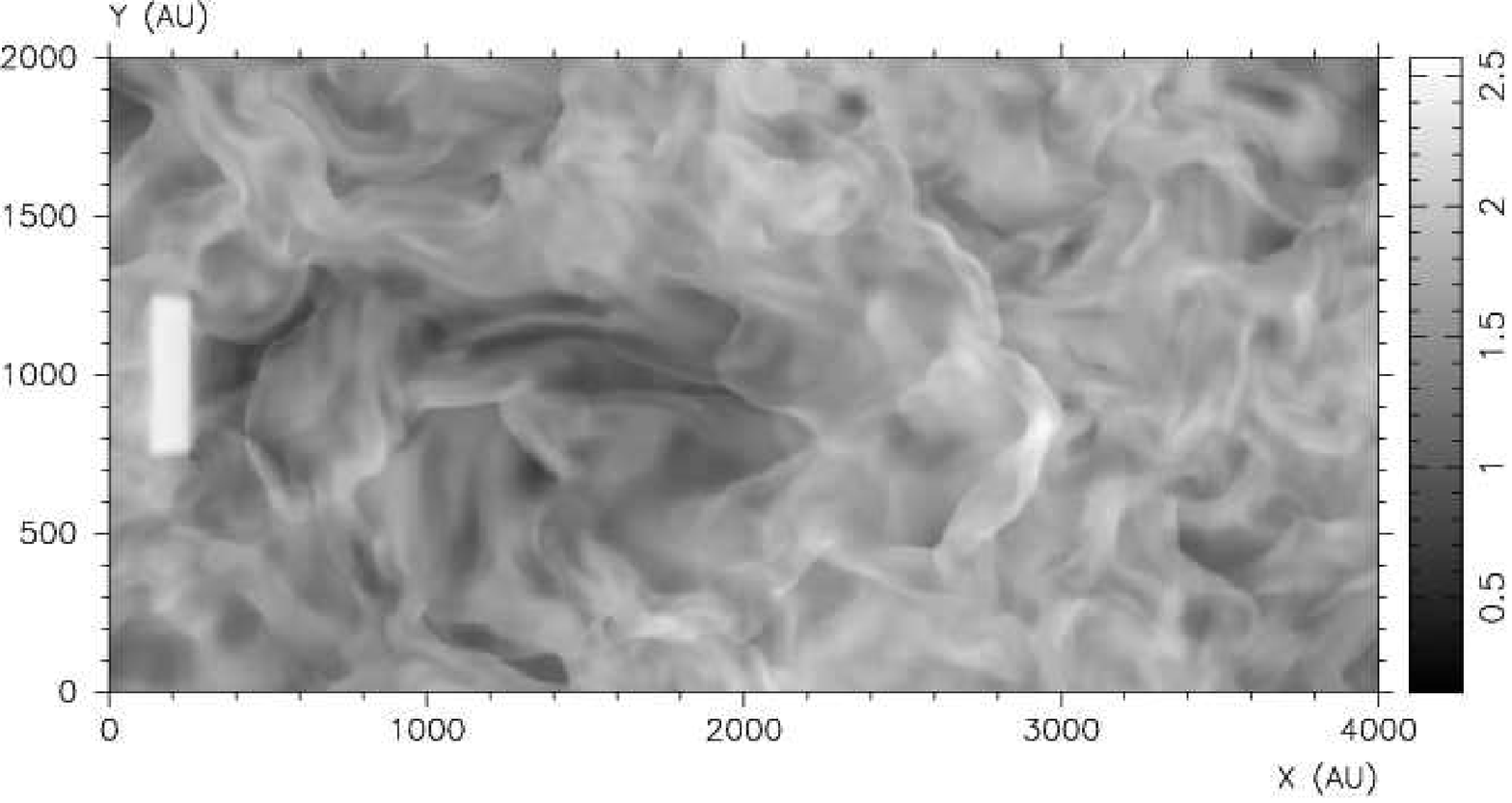}
  \includegraphics[angle=-90,clip=true,width=0.32\textwidth]{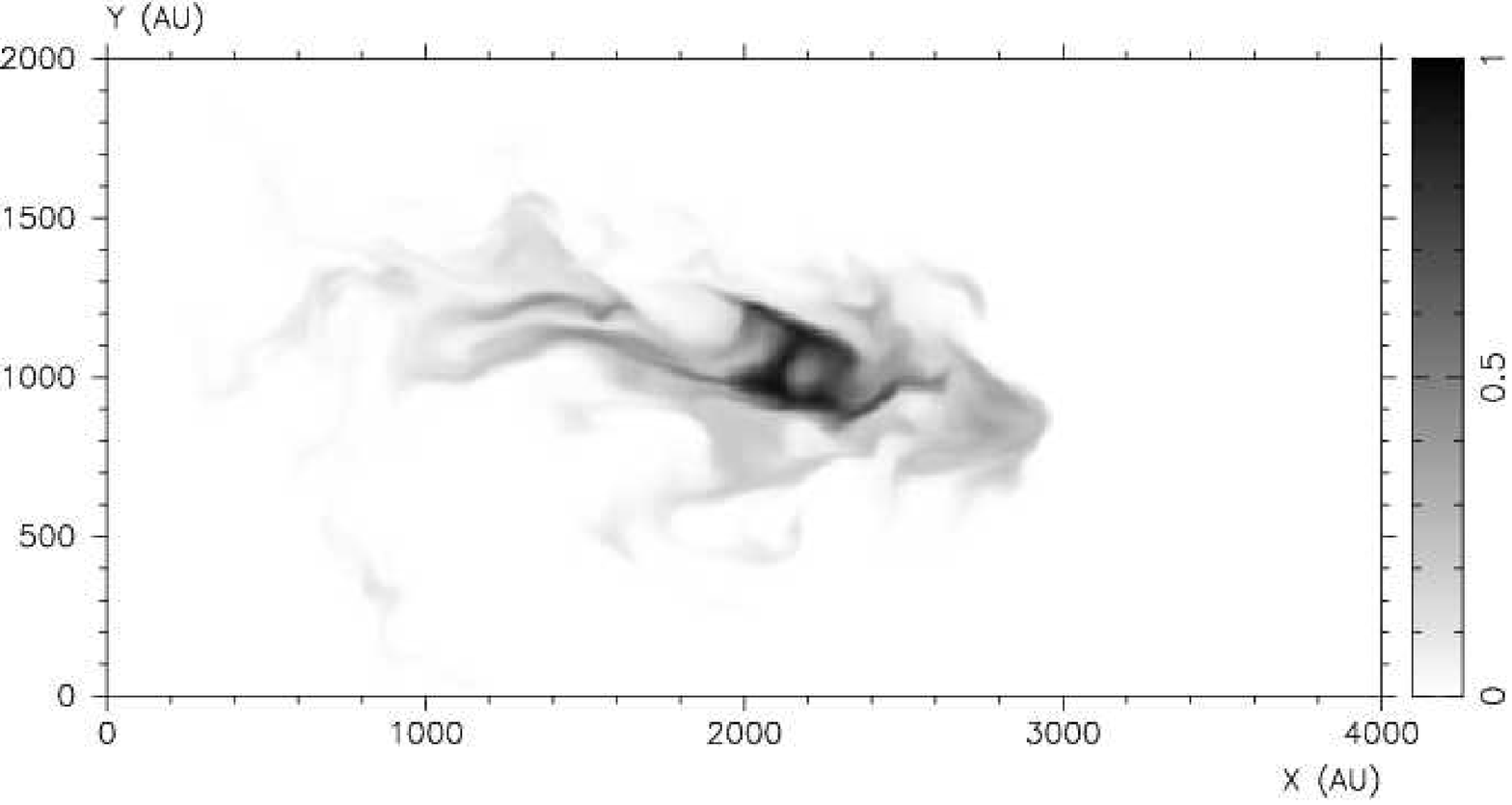} \\
  \includegraphics[angle=-90,clip=true,width=0.32\textwidth]{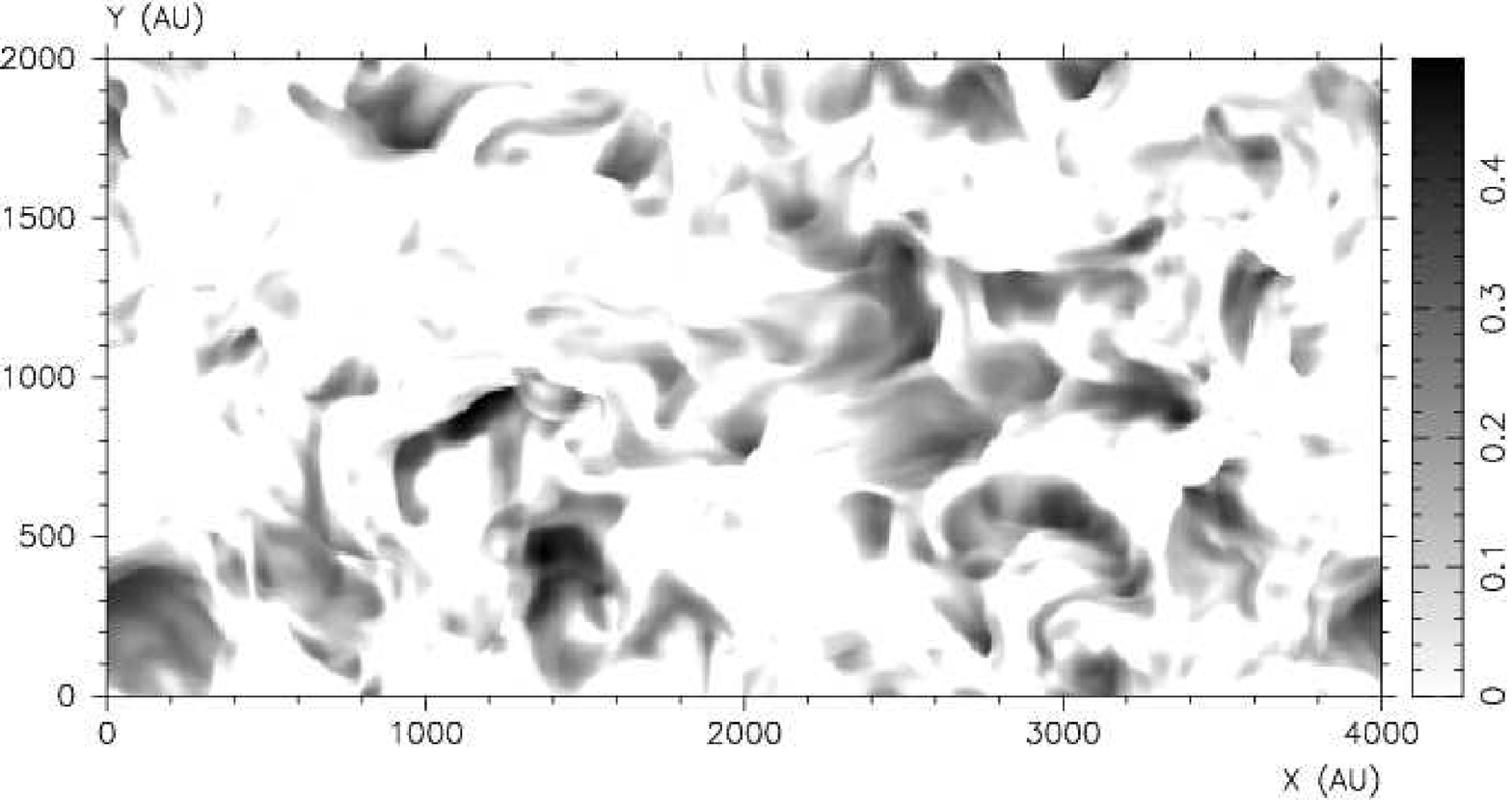}
  \includegraphics[angle=-90,clip=true,width=0.32\textwidth]{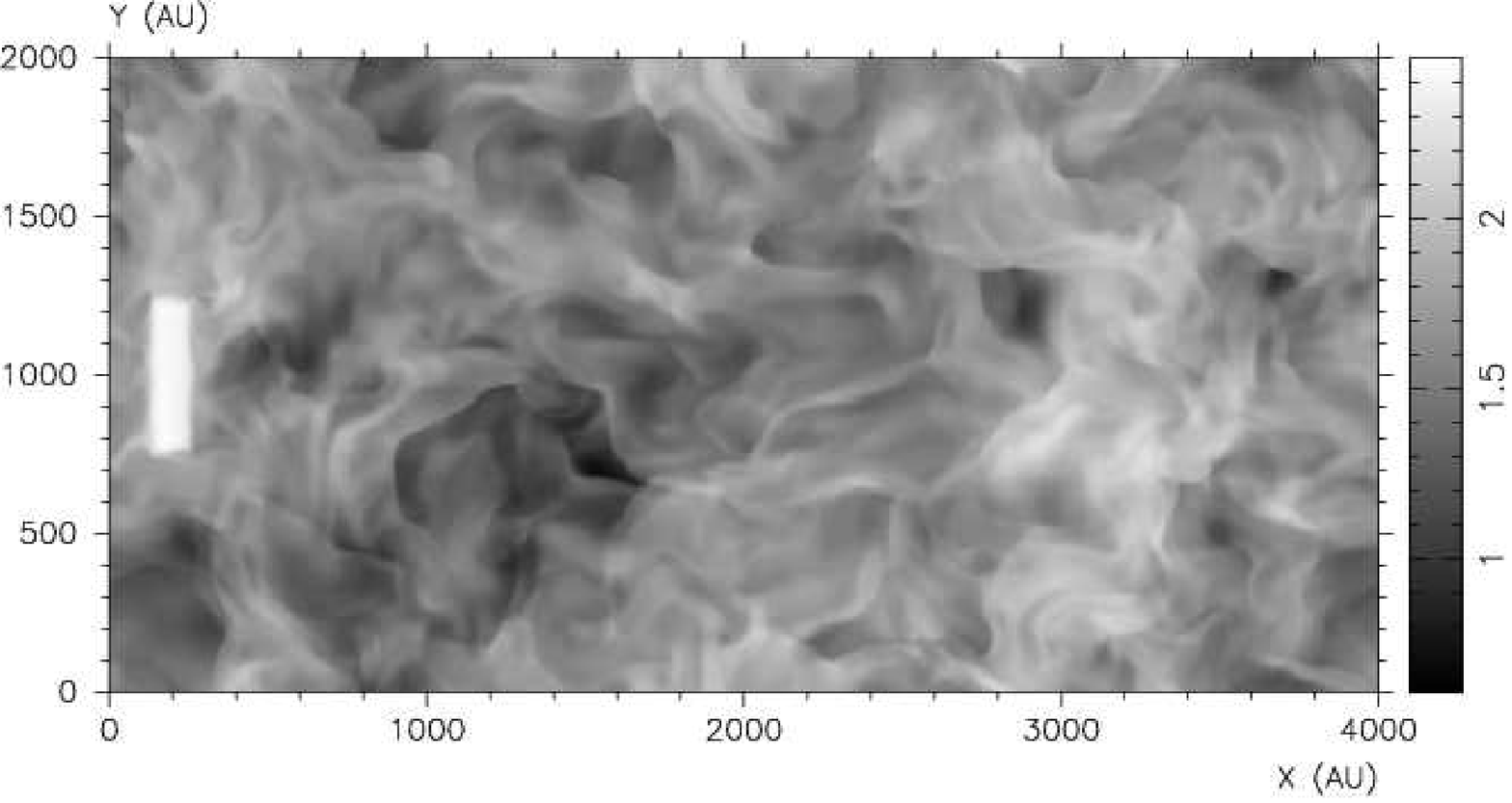}
  \includegraphics[angle=-90,clip=true,width=0.32\textwidth]{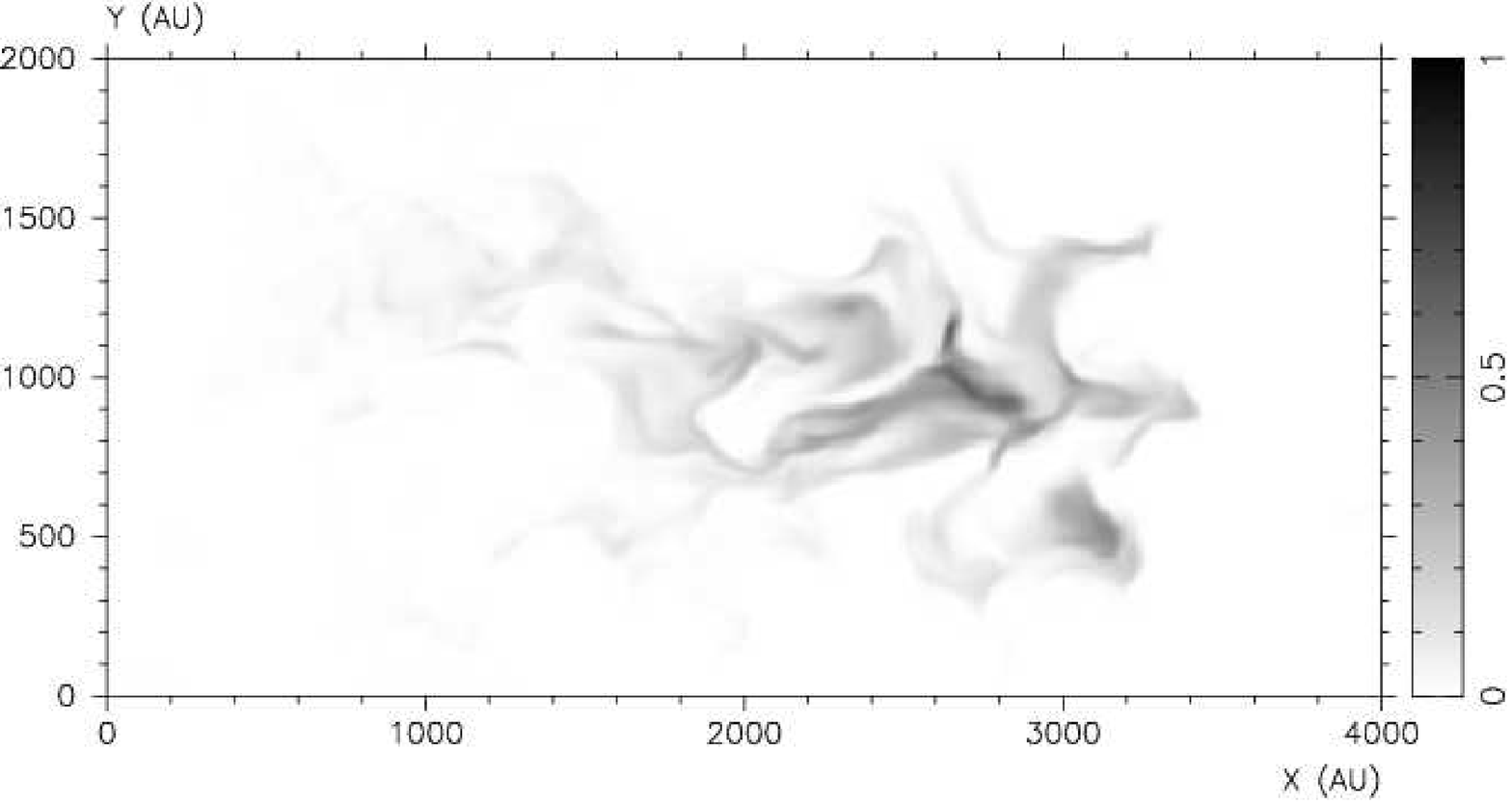} \\
  \includegraphics[angle=-90,clip=true,width=0.32\textwidth]{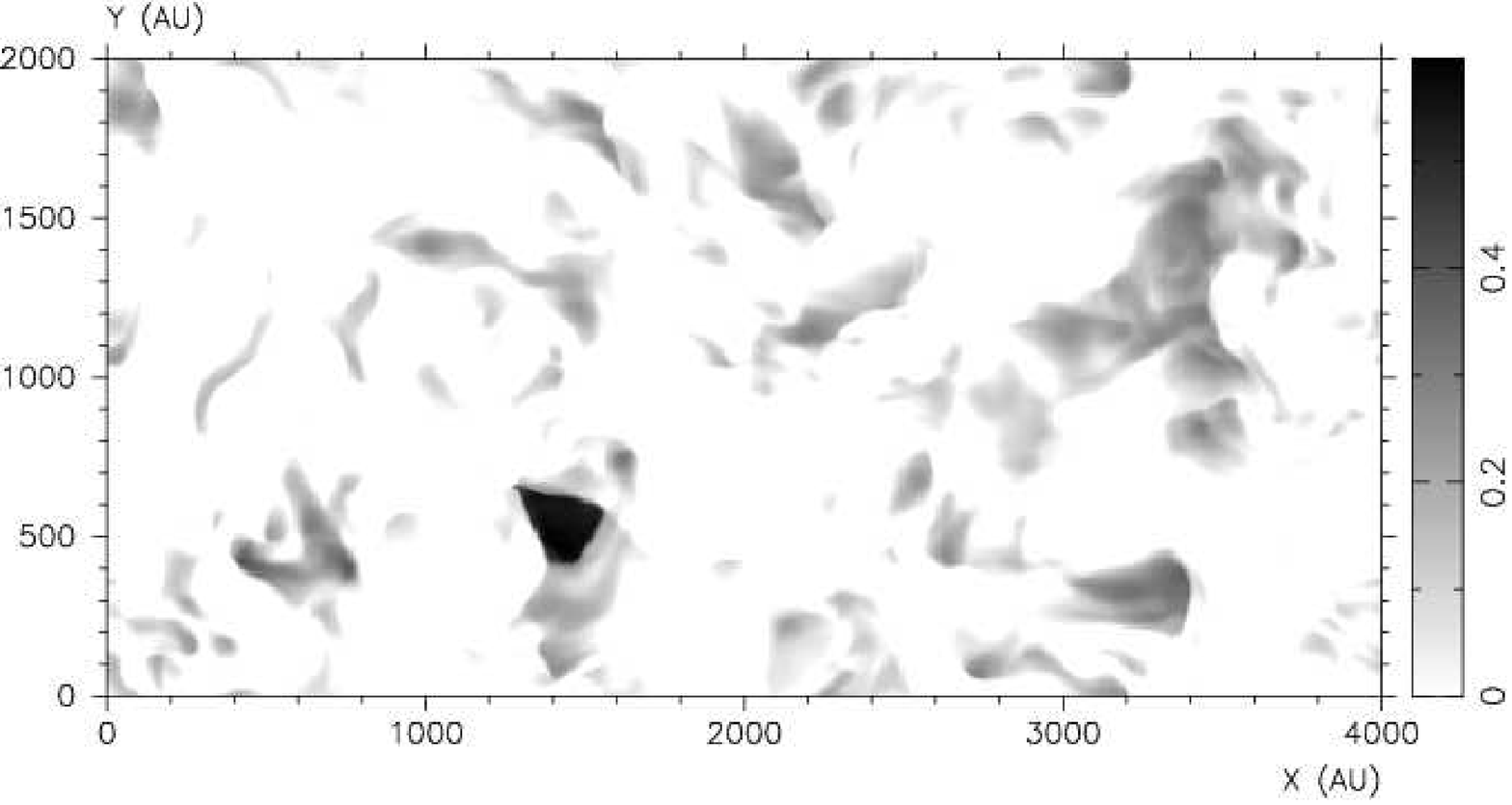}
  \includegraphics[angle=-90,clip=true,width=0.32\textwidth]{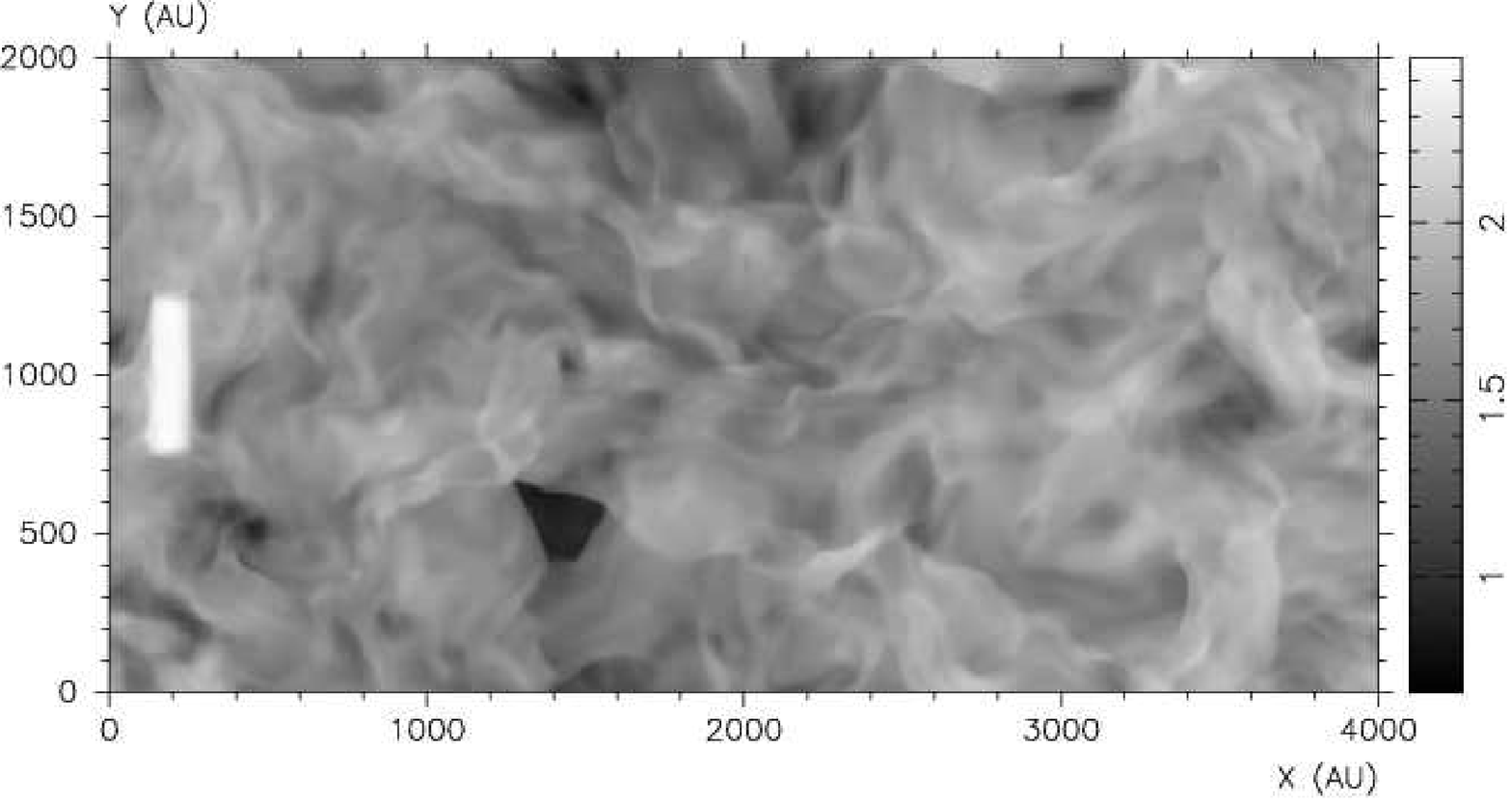}
  \includegraphics[angle=-90,clip=true,width=0.32\textwidth]{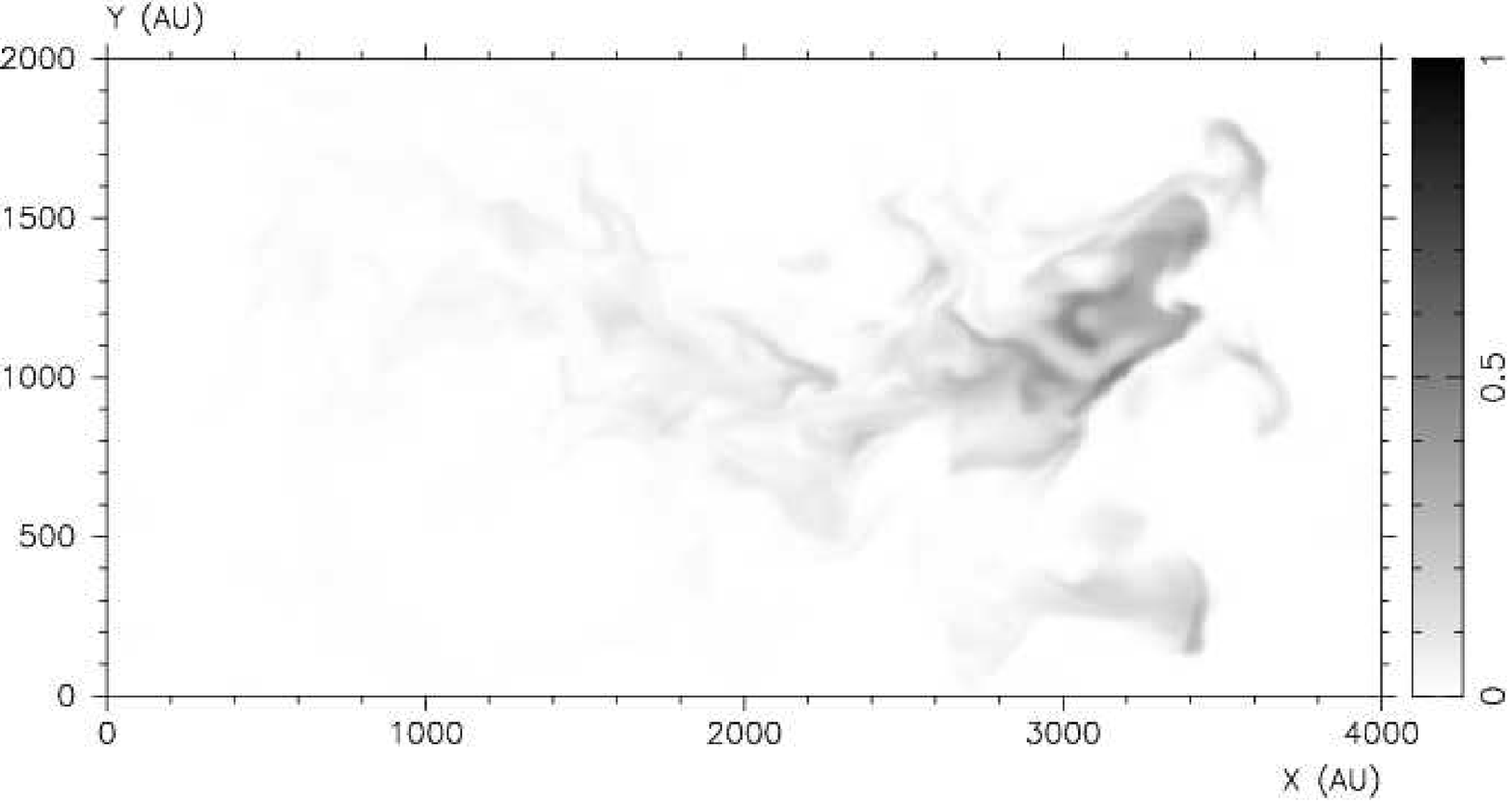} \\
\end{center}
\caption{Evolution of the short-lived jet simulation, run~1 showing
cross cuts about the simulation domain mid-plane of the logarithm of
the Mach number (left), the logarithm of the gas density in
$\textrm{cm}^{-3}$, (center) and the spatial distribution of the
fraction of outflow ejected gas as followed by an advected contaminant
(right) at $t=92,~230,~461,~691,~\textrm{and}~922~\textrm{yr}$ from
top to bottom. By the end of the simulation the jet-launched gas is
widely distributed throughout the domain, the outflow cavity has been
thoroughly subsumed into the ambient turbulence and much of the domain
retains supersonic flow speed.  Furthermore, the flows with the
highest Mach number at the end of the simulation correspond to regions
containing the highest concentration of jet gas.\label{f4}}
\end{figure}

The time dependent evolution of the outflow cavity driven by the short
jet pulse is illustrated in figure \ref{f4}.  The figure is composed
of several plots which show cross cuts about the simulation mid-plane
of the logarithm of the Mach number (left), the logarithm of the gas
density in $\textrm{cm}^{-3}$ (center), and the spatial distribution
of the fraction of outflow ejected gas as followed by an advected
contaminant (right) at
$t=92,~230,~461,~691,~\textrm{and}~922~\textrm{yr}$ from top to
bottom.  The color mapping of the plots of Mach number is constructed
to reveal regions of supersonic flow only.  The images in the right
column showing the advected ``jet contaminant'' reveal that by the end
of the simulation the ejected gas has been widely distributed by the
turbulent flow across the computational domain.  The end of the
simulation at $t=922~\textrm{yr}$ corresponds to approximately one
turbulent crossing time across the shorter edges of the computational
domain of $L/v_{turb} \sim 200~\textrm{AU}/10~\textrm{km s}^{-1} =
948~\textrm{yr}$.  By approximately one bow-shock radius crossing
time, $t \sim 461~\textrm{to}~691~\textrm{yr}$, the initially laminar
jet flow has evolved into a turbulent flow indistinguishable from the
turbulent motions of the cloud.  Furthermore, the flows with the
highest Mach number at the end of the simulation correspond to regions
containing the highest concentration of jet gas.  This indicates that
one crossing time after the outflow source expires the transform of
bulk, directed fossil cavity motion into a fully turbulent flow is
complete.
\subsection{Protostellar Jets and the Production of Turbulence}
The question of if, where and how protostellar outflows drive
turbulence in their environments remains an open issue.  In this work
we have attempted to address one facet of the problem.  In the recent
work of \cite{banerjee} the problem of outflow driven turbulence was
also addressed however these authors came to a very different
conclusion then we have in this and previous works.  In
\cite{banerjee} single outflow cavity structures where used as a means
of determining the efficacy of outflows as a source of supersonic
turbulence.  In that work it was argued that regardless of the
momentum imparted to the cloud by an outflow, effective
outflow-turbulence coupling will only occur if a significant volume
fraction of the outflow cavity retains supersonic flow speed at the
time it is disrupted into the cloud.  Using numerical simulations
\cite{banerjee} found that most of the volume overrun by either
transient or continuously driven isothermal jet models remain subsonic
after passing trough the bow-shock and thereby conclude that outflows
are unlikely to act as a significant source of supersonic turbulence
in molecular clouds.  The results in our paper point towards the
opposite conclusion. In what follows we discuss the differences
between the basic scenario for jet driven turbulence presented in
\cite{banerjee} and the ideas which underpin this work and our
previous studies \cite{cunningham-jc, quillen, cunningham-cavity}.  In
addition we address specific issues raised by the simulations in
\cite{banerjee}

The basic premise of our work is that fossil cavities driven by transient
protostellar outflows couple strongly to their surrounding cloud
environments and drive turbulence within those clouds.  We conjecture
that the coupling can occur in two ways.  First to produce turbulence
in an initially quiescent medium interactions in the form of
collisions between fossil cavities are required.  Second, once fully
developed turbulence exists individual fossil outflows can still give
their energy and momenta back to the turbulent flow through the
disruption of the cavity shell.  The simulations presented here
address the second process.  With regard to the initial generation of
turbulence via interacting fossil cavities we note a few key points.
First, as shown in \cite{cunningham-jc}, in a typical cluster
environment one can expect that every parcel of ambient gas will be
overrun by an outflow at least once.  One can also estimate occurrence
of collisions.  For clouds containing a volumetric density of $N_*$
protostellar driven bipolar outflows each of radius $R$, length $L$
and lifetime $t_{out}$ the probability P of cavity overlap increases
as:
\begin{equation}
P = N_*^2\pi R^2 L \frac{t_{out}}{t_{cloud}}
\end{equation}
where $t_{cloud}$ is the age of the cloud which we assume fixed.  Thus
the older the outflow the higher the probability that cavities will
overlap.  In the work of \cite{cunningham-jc} direct numerical
simulations showed that active outflows, where jet beams collided,
were less efficient at stirring ambient material than the sum of two
similar outflows which do not collide.  This was because the beams
could not pass through each other and forced a shock mediated
redirection in which significant kinetic energy of the flow was lost
to radiation.  The conclusion of that study was not, however, that
collisions of all kinds where ineffective. Instead the authors
conjectured that interactions between slower, longer lived (and
larger) fossil cavities could be the source of turbulent driving.
Indeed the important difference between active outflows and fossil
outflow cavities is in the later case overlapping outflows can
penetrate through each other causing disruption but not redirection.
This is due to the fact that the fossil cavities are hollow and
contain no active momentum bearing jet beam.  We note that the work of
\cite{matzner} provides an analytical description of just such a
process. Using dimensional arguments \cite{matzner} derives a length
scale at which impulsively driven (spherical) outflows will interact.
Given an ambient cloud of density $\rho$ and volumetric outflow rate
$S$ and with each outflow driven with momenta $I$ Matzner finds
turbulence will be driven at a scale length:
\begin{equation}
L_{turb} \sim \frac{I^{1/7}}{\rho^{1/7} S^{1/7}}
\end{equation}
Typical values of $\rho$, $I$ and $S$ appropriate for NGC 1333 yield
an interaction scale $L_{turb} \sim 1~\textrm{pc}$.  Simulation
studies presented in \cite{linaka} and \cite{nakamura}support the
basic scenario that multiple outflows can drive turbulence in a
protstellar cloud.  In addition \cite{carroll} have directly confirmed
Matzner's prediction with a a longer effective scale length
obtained due to jet collimation of $L_{eff} \approx 3 L_{turb}$.  Thus
we conclude that multiple overlapping, interacting fossil cavities can
drive turbulence in an initially quiescent medium.

In this paper we have shown that interactions need not occur for a
fossil cavity, already in a turbulent medium, to resupply its
environment with turbulent energy.  Our identification of the
disruption and eventual subsumption of the fossil cavities
demonstrates the pathways through which an outflow interacting with
only ambient turbulence can act to support a turbulent cascade.  This
can been seen in the time evolution of the Mach number of the short
jet pulse driven outflow cavity (run~1) shown in the left column of
figure \ref{f4}, revealing that the outflow cavity becomes
significantly disrupted and subsumed into the ambient turbulence
before becoming subsonic.  Thus in the present simulations, the
seeding of flow instability via ambient turbulence provides the
mechanism for disruption.

The scenario through which protostellar outflows drive turbulence
proposed and studied in this work contrasts with that of
\cite{banerjee}.  The primary reason for the difference is that the
two works strive to answer different questions.  Whereas
\cite{banerjee} seek to determine the efficacy of isolated outflows in
a quiescent media, the present work focuses on the efficacy of
outflows toward sustaining an initial state of supersonic turbulence.
Our scenario may also be contrasted with the assumptions of
\cite{banerjee} who began their simulations with a single isothermal
jet propagating into a quiescent ambient medium.  They seek to test if
a single jet can leave behind enough supersonic material to power
turbulence on its own.  Models arising from an initially turbulent
state should be viewed as more realistic as clouds are never
quiescent.  Molecular clouds will be always be born in an environment
characterized by a range of motions on different size scale due to
formation: i.e. either gravitational collapse or collisions between
large scale flows.  In this regard we recall that both observational
and theoretical studies of molecular outflows show mass-velocity
relations which scale as $M(v) \propto v^{k}$ with $=1.3<k<2.5$
\citep{lada,smith,lee}.  Thus while the bulk of material will be
moving at low velocity much of the gas set in motion is supersonic.
This is because the ambient environment is laced with a volume filling
supersonic flow which will form a ``floor'' or background into which
the outflow propagates. Thus the assumptions underlying the claims
that jets can not drive turbulence should be considered in light of
the environments these jets inhabit.  Such consideration as shown here
and in the work of \cite{nakamura} and \cite{carroll} shows that jets
can be quite effective in driving and maintaining turbulent cascades.

We note also other factors which may cause lower estimates of flow
speeds and momentum imparted to entrained gas in the models of
\cite{banerjee}.  The outflows presented in this paper are driven at
the high Mach number, $M_j \approx 90$ relative to the sound speed of
the ambient gas, which is to be expected in YSO jets.  The
continuously driven jet models in this paper achieve a characteristic
bow-shock compression ratio of $150$, consistent with the analytic
estimate given by equation 5 of \cite{blondin}.  In contrast, the
simulations of \cite{banerjee} consider isothermal hydrodynamic jets
driven by flows with Mach numbers, $M_j$ of $5$ and $10$ (with a
single $M_j = 20$ case interacting with a cloud).  Because the $M_j =
5 \& 10$ jet models are driven with density contrast $\rho_j/\rho_a =
1$, the inward and outward facing shocks at the jet Mach disk
propagate at half the driving speed of the jet with shock Mach number
$M_s$ of $2.5$ and $5$.  Consideration of Rankine-Hugoniot jumps
across a one-dimensional shock yields the prediction that these models
would achieve compression ratios of $M_s^2$, or $6.25$ and $25$,
consistent with the density plots presented in that study. Thus lower
compression ratios are achieved than is expected from higher Mach
number jet models.  Higher Mach number models exhibit more highly
compressed flow structures take longer to decelerate to a subsonic
regimes.

Our models also consider the effects of radiative line cooling which
is more characteristic of actual protostellar outflow environments.
The simulations in \cite{banerjee} utilize nearly isothermal flow
conditions via a choice of the polytropic index $\gamma = 1.0001$.
While useful for approximating the compressive effects of cooling, it
is important to note the limitations of such an approach which can, in
general, produce differences from a conditions in which cooling is
applied.  While the isothermal approximation will produce high
compressions near shocks which are characteristic of flows subject to
rapid radiative energy loss, it does not capture the effect of
mechanical cooling in rarefactions.  Such effects are particularly
important for transient outflow models which employ a time-dependent
jet launching. This effect can be illustrated by considering the one
dimensional impulsive withdrawal of a piston with constant speed
$v_p$, from a uniform gas with sound speed $c_o$.  The leading front
of the rarefaction has sound speed $c$ given as,
\[
c = \min\left[c_o - \frac{\gamma-1}{2}|v_p|,0\right].
\]
This simplifies to $c=c_o$ for models which use an isothermal equation
of state.  While isothermal models adequately approximate the effect
of rapid radiative energy loss and high compression across strong
shocks, such models do not capture the effect of mechanical cooling
across rarefactions.

Strong rarefactions which ensue after the driving source of the
outflow expires are characteristic of outflow cavity/transient outflow
models \cite{cunningham-cavity}.  Because post-rarefaction isothermal
flows are not subject to the effect of mechanical cooling, the sound
speed in these flow regions is overestimated and will likely causes a
systematic underestimation of the Mach number in such regions of the
flow field.  This effect biases the results of such models in a manner
that is adverse to the development supersonic motions.  These effects
should be studied in future works to ascertain their impact on the
volumetric distribution of mach numbers in single jets driven into a
quiescent medium.
\section{Conclusion}
The results presented in this paper show that outflow driven cavities
contribute to turbulence in their immediate environment provided there
exists some mechanism to disrupt the (mostly) laminar flow patterns
that would result from an undisturbed outflow in a quiescent
environment.  The particular form of the disruption in this work is
that of decaying turbulence in the ambient environment with an initial
turbulent velocity that is nominally 10\% that of the propagation
speed of the outflow bow-shock.  We have shown that even structures
which are denser and propagate more quickly than the ambient
turbulence are significantly disrupted, despite their higher momentum
density.  Outflow structures can be disrupted by the development of
instabilities such as such as radiative \citep{sutherland} and thin
shell \cite{vishniac1,vishniac2} modes which have growth times far
shorter then the evolutionary timescales of the flow.  The wide
spectrum of motions present in the ambient turbulence provides a
multi-mode spectrum of perturbation seeds which initiate the growth of
these instabilities.  Extinct outflow cavities thus become fully
subsumed into the background turbulence in the time required for these
disruptions to cross the cavity.  Observation of outflow driven
cavities similar to the work carried out in \cite{quillen} therefore
become increasingly confused with the ambient turbulence during this
transition.  For the case of highly collimated outflow structures
which are prototypical of low-mass star formation the relevant
crossing distance is the radius of the roughly cylindrical outflow
bow-shock.  The disruption time for low mass protostellar outflows is
therefore short relative to the turbulent crossing time of their
parent cores.  After the disruption time, the outflow cavities act to
support turbulence in the local environment on a driving scale
comparable to the size of the outflow cavity.  We conclude therefore
that protostellar outflows provide a efficient form of dynamical
feedback in low mass star forming cores as suggested by
\cite{normansilk}.

We postulate that the action which disrupts an outflow cavity need not
be fully developed turbulence in the ambient environment.  The
interaction of an extinct outflow cavity with another outflow or
outflow remnant would produce similar results.  The result of
\cite{cunningham-jc}showed that large radiative energy loss from
the interaction of multiple {\it active} outflow actually inhibits
mechanical support. Here we emphasize cavity-cavity interactions can
also drive turbulence and this mechanism as shown by \cite{carroll}.
Future work should also focus on ecological studies of multiple star
formation \citep{linaka, nakamura}. Simulation of such systems over
long timescales with sufficient resolution could determine the rate of
steady-state turbulent decay resulting from a steady rate of outflow
formation.

\acknowledgments

We wish to thank Chris Matzner and Chris McKee for extremely useful
discussions. Hector Arce, John Bally, Pat Hartigan were also generous
with their time.  Tim Dennis, Kris Yirak, Brandon Schroyer and Mike
Laski provided invaluable support and help.

Support for this work was in part provided by by NASA through awards
issued by JPL/Caltech through Spitzer program 20269, the National
Science Foundation through grants AST-0406823, AST-0507519 and
PHY-0552695 as well as the Space Telescope Science Institute through
grants HST-AR-10972, HST-AR-11250, HST-AR-11252 to.  Andrew Cunningham
received support under the auspices of the US Department of Energy by
Lawrence Livermore National Laboratory under contact
DE-AC52-07NA27344.  We also thank the University of Rochester
Laboratory for Laser Energetics and funds received through the DOE
Cooperative Agreement No. DE-FC03-02NA00057.


\begin{thebibliography}{}
\bibitem[Bally et al.(1996)]{bally96}
\bibitem[Bally \& Reipurth(2001)]{ballyreview} Bally, J, \& Reipurth,
B. 2001, ARAA, 39, 403
\bibitem[Bally et al.(2002)]{bally} Bally, J., Heathcote, S., 
Reipurth, B., Morse, J., Hartigan, P., \& Schwartz, R.\ 2002, \aj, 123, 
2627 
\bibitem[Bally et al.(2006)]{2006AJ....131..473B} Bally, J., Licht, D., 
Smith, N., \& Walawender, J.\ 2006, \aj, 131, 473 
\bibitem[Bally \& Devine(1994)]{1994ApJ...428L..65B} Bally, J., \&
Devine, D.\ 1994, \apjl, 428, L65
\bibitem[Bally et al.(1996)]{1996ApJ...473..921B} Bally, J., Devine, D., 
\& Alten, V.\ 1996, \apj, 473, 921 
\bibitem[Bally \& Reipurth(2003)]{2003AJ....126..893B} Bally, J., \&
Reipurth, B.\ 2003, \aj, 126, 893
\bibitem[Banerjee et al.(2007)]{banerjee} Banerjee, R., Klessen, 
R.~S., \& Fendt, C.\ 2007, \apj, 668, 1028 
\bibitem[Blitz et al.(2007)]{blitz} Blitz, L., Fukui, Y., 
Kawamura, A., Leroy, A., Mizuno, N., \& Rosolowsky, E.\ 2007, Protostars 
and Planets V, 81 
\bibitem[Blondin et al.(1990)]{blondin} Blondin, J.~M., 
Fryxell, B.~A., \& Konigl, A.\ 1990, \apj, 360, 370 
\bibitem[Cabrit(2007)]{2007IAUS..243..203C} Cabrit, S.\ 2007, IAU 
Symposium, 243, 203
\bibitem[Carroll et al.(2008)]{carroll} Carroll, J., et al. \ 2008, in prep
\bibitem[Cunningham et al.(2006a)]{cunningham-jc} Cunningham, A.~J., 
Frank, A., \& Blackman, E.~G.\ 2006, \apj, 646, 1059
\bibitem[Cunningham et al.(2005)]{cunningham-I} Cunningham, A., 
Frank, A., \& Hartmann, L.\ 2005, \apj, 631, 1010 
\bibitem[Cunningham et al.(2006b)]{cunningham-cavity} Cunningham, A.~J., 
Frank, A., Quillen, A.~C., \& Blackman, E.~G.\ 2006, \apj, 653, 416 
\bibitem[Cunningham et al.(2007)]{cunningham-mhd} Cunningham, A.~J., 
Frank, A., Varniere, P., Mitran, S., \& Jones, T.~W.\ 2007, ArXiv e-prints, 
710, arXiv:0710.0424 
\bibitem[Delamarter et al.(2000)]{delamarter} Delamarter, G., 
Frank, A., \& Hartmann, L.\ 2000, \apj, 530, 923 
\bibitem[Elmegreen(2000)]{elmegreen} Elmegreen, B.~G.\ 2000, 
\apj, 530, 277 
\bibitem[Field et al.(2008)]{field} Field, G.~B., Blackman, 
E.~G., \& Keto, E.~R.\ 2008, \mnras, 385, 181 
\bibitem[Fleck(1980)]{fleck} Fleck, R.~C., Jr.\ 1980, \apj, 
242, 1019 
\bibitem[Goldreich \& Kwan(1974)]{goldreich} Goldreich, P., \& 
Kwan, J.\ 1974, \apj, 189, 441 
\bibitem[Hueckstaedt et al.(2006)]{hueckstaedt} Hueckstaedt, R.~M., 
Hunter, J.~H., \& Lovelace, R.~V.~E.\ 2006, \mnras, 369, 1143 
\bibitem[Klein et al.(2007)]{klein} Klein, R.~I., Inutsuka, 
S.-I., Padoan, P., \& Tomisaka, K.\ 2007, Protostars and Planets V, 99 
\bibitem[Knee \& Sandell(2000)]{knee} Knee, L.~B.~G., \& 
Sandell, G.\ 2000, \aap, 361, 671
\bibitem[Krumholz \& McKee(2005)]{krumholz-tr} Krumholz, M.~R., \& 
McKee, C.~F.\ 2005, \apj, 630, 250 
\bibitem[Krumholz et al.(2005)]{krumholz05} Krumholz, M.~R., 
McKee, C.~F., \& Klein, R.~I.\ 2005, \apjl, 618, L33 
\bibitem[Krumholz et al.(2006)]{krumholz} Krumholz, M.~R., 
Matzner, C.~D., \& McKee, C.~F.\ 2006, \apj, 653, 361 
\bibitem[Hartigan et al.(2005)]{hartigan} Hartigan, P., 
Heathcote, S., Morse, J.~A., Reipurth, B., \& Bally, J.\ 2005, \aj, 130, 
2197 
\bibitem[Heyer \& Brunt(2004)]{heyer} Heyer, M.~H., \& Brunt, 
C.~M.\ 2004, \apjl, 615, L45 
\bibitem[Lada \& Fich(1996)]{lada} Lada, C.~J., \& Fich, M.\ 1996, \apj, 459, 638 
\bibitem[Larson(1981)]{larson} Larson, R.~B.\ 1981, \mnras, 
194, 809 
\bibitem[Lebedev et al.(2004)]{lebedev} Lebedev, S.~V., et al.\ 
2004, \apj, 616, 988 
\bibitem[Lee et al.(2001)]{lee} Lee, C.-F., Stone, J.~M., 
Ostriker, E.~C., \& Mundy, L.~G.\ 2001, \apj, 557, 429 
\bibitem[Li \& Nakamura(2006)]{linaka} Li, Z.-Y., \& Nakamura, 
F.\ 2006, \apjl, 640, L187 
\bibitem[Mac Low(1999)]{maclow-dissipation} Mac Low, M.-M.\ 1999, \apj, 
524, 169 
\bibitem[Mac Low(2003)]{maclow} Mac Low, M.-M.\ 2003, 
Turbulence and Magnetic Fields in Astrophysics, 614, 182
\bibitem[Mac Low(2000)]{maclow-outflows} Mac Low, M.-M.\ 2000, Star 
Formation from the Small to the Large Scale, 445, 457 
\bibitem[Mac Low \& Klessen(2004)]{maclowkl} Mac Low, M.-M., \& 
Klessen, R.~S.\ 2004, Reviews of Modern Physics, 76, 125 
\bibitem[McGroarty et al.(2004)]{McGroarty} McGroarty, F., Ray, T.~P., \& Bally, J.\ 2004, \aap, 415, 189
\bibitem[Matzner(2001)]{matzner01} Matzner, C.~D.\ 2001, From 
Darkness to Light: Origin and Evolution of Young Stellar Clusters, 243, 757 
\bibitem[Matzner \& McKee(2000)]{matzner00} Matzner, C.~D., \& 
McKee, C.~F.\ 2000, Bulletin of the American Astronomical Society, 32, 883 
\bibitem[Matzner(2007)]{matzner} Matzner, C.~D.\ 2007, \apj, 
659, 1394 
\bibitem[Nakamura \& Li(2007)]{nakamura} Nakamura, F., \& Li, 
Z.-Y.\ 2007, \apj, 662, 395 
\bibitem[Norman \& Silk(1980)]{normansilk} Norman, C., \& Silk, 
J.\ 1980, \apj, 238, 158 
\bibitem[Porter et al.(1992)]{porter} Porter, D.~H., Pouquet, 
A., \& Woodward, P.~R.\ 1992, Physical Review Letters, 68, 3156
\bibitem[Quillen et al.(2005)]{quillen} Quillen, A.~C., 
Thorndike, S.~L., Cunningham, A., Frank, A., Gutermuth, R.~A., Blackman, 
E.~G., Pipher, J.~L., \& Ridge, N.\ 2005, \apj, 632, 941
\bibitem[Stanke et al.(1999)]{1999A&A...350L..43S} Stanke, T.,
McCaughrean, M.~J., \& Zinnecker, H.\ 1999, \aap, 350, L43
\bibitem[Shepherd(2003)]{shepherd} Shepherd, D.\ 2003, Galactic 
Star Formation Across the Stellar Mass Spectrum, 287, 333 
\bibitem[Smith et 
al.(1997)]{smith} Smith, M.~D., Suttner, G., \& Yorke, H.~W.\ 1997, \aap, 323, 223 
\bibitem[Stone et al.(1998)]{stone} Stone, J.~M., Ostriker, 
E.~C., \& Gammie, C.~F.\ 1998, \apjl, 508, L99 
\bibitem[Sutherland et al.(2003)]{sutherland} Sutherland, R.~S., 
Bicknell, G.~V., \& Dopita, M.~A.\ 2003, \apj, 591, 238 
\bibitem[Vishniac(1994)]{vishniac2} Vishniac, E.~T.\ 1994, \apj, 
428, 186 
\bibitem[Vishniac \& Ryu(1989)]{vishniac1} Vishniac, E.~T., \& 
Ryu, D.\ 1989, \apj, 337, 917 
\bibitem[Warin et al.(1996)]{warin} Warin, S., Castets, A., 
Langer, W.~D., Wilson, R.~W., \& Pagani, L.\ 1996, \aap, 306, 935 
\bibitem[Whelan et al.(2005)]{whelan} Whelan, E.~T., Ray, 
T.~P., Bacciotti, F., Natta, A., Testi, L., \& Randich, S.\ 2005, \nat, 
435, 652 
\bibitem[Yirak et al.(2007)]{yirak} Yirak, K., Frank, A., 
Cunningham, A., \& Mitran, S.\ 2007, ArXiv e-prints, 705, arXiv:0705.1558 
\end{thebibliography}
\end{document}